\documentclass[journal,10pt]{IEEEtran} 
\usepackage{balance}
\usepackage{algpseudocode}
\usepackage[ruled,vlined,linesnumbered,resetcount]{algorithm2e}
\SetKwRepeat{Do}{do}{while}%
\usepackage[linesnumbered,ruled,vlined]{algorithm2e}
\usepackage{cite}
\usepackage{amsmath,amssymb,amsfonts,bm}
\usepackage{graphicx}
\usepackage{textcomp}
\usepackage{graphics}
\usepackage{graphicx} 
\usepackage{epsfig}
\usepackage{amssymb}
\usepackage{amsthm}
\usepackage{times} 
\usepackage{amsmath} 
\usepackage{amssymb}  
\usepackage{graphics}
\usepackage{graphicx}
\usepackage{array}
\usepackage{color}
\usepackage[numbers]{natbib} 
\usepackage{mathtools}
\newcommand{\R}{\mathbb{R}}

\newtheorem* {theorem1}{Theorem~1~\normalfont{\cite{han2021reinforcement}}}
\newtheorem* {theorem2}{Theorem~2}
\newtheorem* {theorem3}{Theorem~3}

\newtheorem {lemma}{Lemma}

\newtheorem* {def1}{Definition~1~\normalfont{\cite{thowsen1983uniform}}}
\newtheorem* {def2}{Definition~2~\normalfont{\cite{boyd2004convex}}}
\newtheorem* {def3}{Definition~3~\normalfont{\cite{han2021reinforcement}}}

\def\BibTeX{{\rm B\kern-.05em{\sc i\kern-.025em b}\kern-.08em
    T\kern-.1667em\lower.7ex\hbox{E}\kern-.125emX}}
\usepackage{url}
\usepackage{tikz} 
\usetikzlibrary{shapes,shapes.geometric,arrows,fit,calc,positioning,automata,}

\renewcommand{\SetKwInOut}[2]{%
	\sbox\algocf@inoutbox{\KwSty{#2}\algocf@typo:}%
	\expandafter\ifx\csname InOutSizeDefined\endcsname\relax
	\newcommand\InOutSizeDefined{}%
	\sbox\algocf@inoutbox{\KwSty{#2}\algocf@typo\textbf{:}~}\setlength{\inoutindent}{\wd\algocf@inoutbox}%
	\else
	\ifdim\wd\algocf@inoutbox>\inoutsize%
	\sbox\algocf@inoutbox{\KwSty{#2}\algocf@typo\textbf{:}~}\setlength{\inoutindent}{\wd\algocf@inoutbox}%
	\fi%
	\fi
	\algocf@newcommand{#1}[1]{%
		\ifthenelse{\boolean{algocf@inoutnumbered}}{\relax}{\everypar={\relax}}%
		{\let\\\algocf@newinout\hangindent=\inoutindent\hangafter=1\KwSty{#2}\algocf@typo\textbf{:}~##1\par}%
		\algocf@linesnumbered
}}
\begin{document}
\title{A Learning Approach for Joint Design of Event-triggered Control and Power-Efficient Resource Allocation}
\author{Atefeh. Termehchi, and Mehdi. Rasti, Senior Member,~IEEE
\thanks{Copyright (c) 2015 IEEE. Personal use of this material is permitted. However, permission to use this material for any other purposes must be obtained from the IEEE by sending a request to pubs-permissions@ieee.org. 
Atefeh. Termehchi and Mehdi. Rasti are with the Department of Computer Engineering,
Amirkabir University of Technology, Tehran, Iran (e-mail: {atefetermehchy, rasti}@aut.ac.ir). }}
\maketitle
\begin{abstract}
 In emerging Industrial Cyber-Physical Systems (ICPSs), the joint design of communication and control sub-systems is essential, as these sub-systems are interconnected. In this paper, we study the joint design problem of an event-triggered control and an energy-efficient resource allocation in a fifth generation (5G) wireless network. We formally state the problem as a multi-objective optimization one, aiming to minimize the number of updates on the actuators' input and the power consumption in the downlink transmission. To address the problem, we propose a model-free hierarchical reinforcement learning approach with uniformly ultimate boundedness stability guarantee that learns four policies simultaneously. These policies contain an update time policy on the actuators' input, a control policy, and energy-efficient sub-carrier and power allocation policies. Our simulation results show that the proposed approach can properly control a simulated ICPS and significantly decrease the number of updates on the actuators' input as well as the downlink power consumption.
 \end{abstract}

\vspace{-10pt}
\section{Introduction}
\label{sec:introduction}
Emerging ICPSs, such as smart grid, smart manufacturing, and smart transportation are spatially distributed and high-dimensional. These systems require high reliability, communication between numerous devices, low latency, power-efficient communication, and high computational load \cite{tranoris2018enabling, zerihun2020effect}. To manage these requirements, 5G and beyond 5G networks present a wide range of services that are classified as 1) enhanced mobile broadband (eMBB), 2) ultra-reliable and low-latency communication (URLLC), and 3) massive machine-type communication (mMTC).
The eMBB, URLLC, and mMTC services provide a high data rate with a moderate latency, a communication with low end-to-end delay, and connecting many devices respectively. Because the type of most communications in control sub-systems is URLLC, 5G network is a good choice for exchanging data between a controller and sensors or actuators.\\  
However, there are some serious challenges to deploying 5G network in ICPSs. Specifically, ICPSs with 5G networks have limited network resources and lack the desired stability and performance guarantees \cite{eisen2019control,liu2020latency}. The performance of the control sub-system is defined as achieving the required dynamics response, which is specified by measures of performance such as a desirable steady-state tracking error.
The stability and performance of a control sub-system may be guaranteed through periodic transmissions with a high data rate. It, however, comes at the cost of a higher packet loss rate due to limited resources in wireless networks \cite{lu2015real}. Furthermore, in many applications of ICPSs, most wireless devices rely on batteries and their battery life may be significantly reduced by the increased transmission rate \cite{funk2021learning}. Consequently, the event-triggered control (ETC) method is proposed, in which the transmission times of the control sub-system are triggered based on a predefined event instead of a periodic transmission. This event is characterized according to the stability and performance requirement of the control sub-system.\\ In recent years, extensive research has concentrated on different classes of ETC strategies; see \cite{ge2019distributed, asurvey} and the references therein. Besides, in this context, there are substantial works, which prove better energy-saving and performance of ETC in comparison with the traditional periodic control \cite{antunes2014rollout, dolk2017event}. Nevertheless, these works analyze only low-dimensional or linear models of control sub-systems \cite{funk2021learning, xu2020remote}. Moreover, the analysis of the event-triggered control becomes too complicated when the volatile properties of wireless communication such as delay, limited resources, packet drops, and unreliable links are considered.\\
The design of the event-triggered control in the presence of unreliable links and packet losses has been recently drawn a lot of attention \cite{zhu2020event, wu2020event, xu2020remote}. However, in addition to packet drops, there exist many other features of wireless communication, such as the delay and limited resources, which make a direct impact on the stability and performance of control sub-systems. To deal with these interconnections between the control and communication sub-systems, the joint design method is taken in ICPSs \cite{xu2020reinforcement, liu2019controller, funk2021learning,mamduhi2020cross}. However, developing an analytical model of all control and network features is a fundamental challenge to this method. This is because the sub-systems are typically high-dimensional and the conditions of radio resources are continuously and randomly changing.\\   
Therefore, researchers have used model-free reinforcement learning (RL) in the joint design of ICPS' sub-systems \cite{demirel2018deepcas, vamvoudakis2018model,funk2021learning, xu2020reinforcement,leong2020deep}. In \cite{demirel2018deepcas}, RL is used for proposing a sensors scheduler while the controller is designed beforehand. The actor-critic RL method is also used in \cite{vamvoudakis2018model} to learn the event-triggered control. In \cite{funk2021learning}, option method of Deep RL (DRL) is used for joint optimization of an event policy and a control policy. The event policy determines when the control input should be transmitted and the control policy determines what the control input value should be sent. Nonetheless, the varying characteristics of the wireless network are not considered in \cite{funk2021learning, vamvoudakis2018model}.  
In \cite{xu2020reinforcement}, RL approach is used to jointly design the sampling rate of the control sub-system and the modulation type of the wireless network.\\ Although stability is an essential property for every control sub-system, RL methods could hardly guarantee the stability and reliability of a learning-based controller \cite{han2020actor}. Nonetheless, in \cite{han2020actor, han2021reinforcement,zhang2021safe}, a learning-based controller with uniformly ultimate boundedness (UUB) stability guarantee is proposed, which can be usefully employed in ICPSs with safety constraints. In general, UUB stability says that if the norm of starting state variables of a control sub-system is less than a specified value, then the state variables will eventually enter the neighborhood of the sub-system's equilibrium within a finite time and will never escape from this neighborhood set afterwards \cite{han2021reinforcement}.\\
The goal of this paper is to jointly design the event-triggered control and the energy-efficient allocation of radio resources in an ICPS. To the best of our knowledge, this joint design problem has not yet been studied. We propose to use a novel Hierarchical RL (HRL) approach with UUB stability guarantee to solve the problem.  
Our contributions are as follows. 
 \begin{itemize} 
    \item We assume an ICPS containing multiple eMBB users and a control plant with multiple URLLC users sharing a single cell Orthogonal Frequency-Division Multiple Access (OFDMA) network. We formulate the joint design of the event-triggered control and the energy-efficient resource allocation in the ICPS as a multi-objective optimization problem. The goals of the problem are both minimizing the number of updates on the actuators' input and the energy consumption in the downlink. The constraints of this problem contain the dynamics and UUB stability of the control plant, the minimum Quality of Service (QoS) demand of eMBB and URLLC users, and the power and sub-carrier constraints of the OFDMA network.   
   \item The problem is high-dimensional, complicated and associated with a hybrid action space. To handle these properties, we combine Cascade Attribute Learning Network (CAN) method and option-critic method to develop a novel model-free HRL approach with
   UUB stability guarantee. First, we use CAN method and decouple the problem into two low-dimensional sub-problems of control and resource allocation. We show that using the decoupling method leads to a Pareto solution to the optimization problem. In the second step, we use option-critic method, which is reformulated as Double Actor-Critic (DAC) architecture, to address each sub-problem with a hybrid action space. 
   \item The novel model-free HRL with
   	UUB stability guarantee can simultaneously learn four policies: 1) update time policy on the actuators' input, 2) control policy, which determines the value of control input, 3) energy-efficient sub-carrier allocation policy, and 4) energy-efficient power allocation policy.
   \item We demonstrate the effectiveness and capability of the proposed approach by several simulation results. In comparison with a disjoint and model-based method, our numerical simulation results show that both the number of updates on the actuators' input and the downlink energy consumption are reduced significantly by applying the proposed approach. Moreover, we show the capability of the proposed approach compared with the soft actor-critic algorithm.  
\end{itemize}
\indent This paper is outlined as follows. The system model and problem formulation is described in Section II. The proposed approach is presented in Section III. In Section IV, simulation results are discussed. In Section V, the paper's conclusion and future work are given.
\begin{figure}[t]
	\centerline{\includegraphics[width=0.5\textwidth]{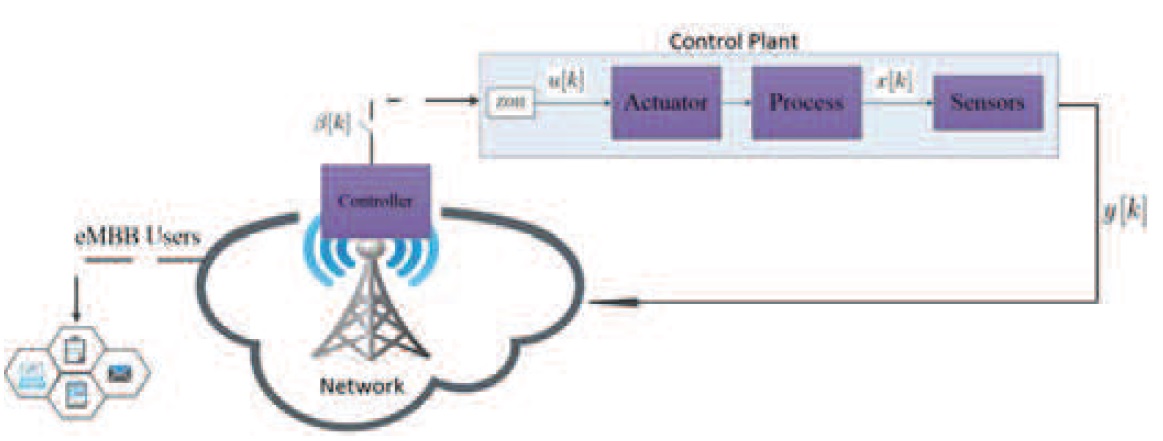}}
	\label{fig:System model}
	\caption{System model of the considered ICPS}
\end{figure} 
\section{SYSTEM MODEL and PROBLEM FORMULATION} 
\subsection{System Model}
 Consider a model of ICPS that consists of 1) a control plant, 2) a central event-triggered controller, and 3) a downlink model of an OFDMA cellular network (Fig. 1). The state values of the control plant, are measured by multiple sensors and sent to the event-triggered learning-based controller. Next, the controller calculates and sends the control input to actuators, whenever required, through the OFDMA single cell network.
 Following the 5G architecture explained by International Telecommunication Union (ITU), the central learning-based controller is supposed to run on a specific or shared hardware in the central office data center layer, which is placed near the network's Base Station (BS) \cite{HUAWEI}.\\
 \textbf{Control Plant}: We suppose dynamics of the control plant is unknown, that is:
 \begin{equation}
 \begin{split}
 \ x[k+1] & = f(x[k], u[k], \omega[k]),\\
   y[k] &= v(x[k]),
 \label{eqn:pl1}
 \end{split}
 \end{equation}
where $f(.)$ and $v(.)$ are unknown functions, $x[k]\in \R^{n}$, $u[k]\in \R^{m}$, and $y[k]\in \R^{q}$ denote the vector of the control state, control input, and sensors' output at discrete time $k\in\mathcal{K}$ ($\mathcal{K}= \{1,2,...,K\}$) respectively. Also, vector $\omega[k]\in \R^{m}$ is actuation disturbances at discrete time $k\in\mathcal{K}$.
We assume the control plant, described by dynamics (\ref{eqn:pl1}), is completely state observable, as it is regularly assumed in the related literature, e.g. \cite{radac2020robust}.\\ 
\textbf{Event-triggered Controller}: When the sensors' output vector ($y[k]$) is received by the central event-triggered controller, it decides whether actuators' input should update ($\beta[k]= 1$) or ignore the update and save wireless resource ($\beta[k]= 0$). This decision has been taken based on UUB stability guarantee of the control plant, defined in what follows.
\begin{def1} \label{D1}
	A control plant is uniformly ultimately bounded with ultimate bound $\rho$, if there are positive constants $b, \rho$ and $\forall \zeta < b:\exists T(\zeta,\rho)$, such that $||x[k_0]||<\zeta	\Rightarrow||x[k]||<\rho, \forall k \geq k_0 + T$. If $\zeta$ can be arbitrary large, then the control plant is globally uniformly ultimately bounded.	
\end{def1}
\noindent In addition, if update variable $\beta[k]= 1$, then the controller calculates the control input variable ($u[k]$) considering UUB stability. We assume that Zero Order Hold (ZOH) holds actuators' input constant between two consecutive updates. This can be mathematically given by:
\begin{equation}
\ u[k]  = u[k] \beta[k] + u[k-1](1- \beta[k]): \beta[k] \in \{0,1\}.
\label{eqn:event-control}
\end{equation}
\textbf{OFDMA Network}:
 We assume the downlink model of a single cell OFDMA network with one BS. The model has $N$ downlink users denoted by $\mathcal{N}=\{1,2,...,N\}$. The downlink users have a set of $N^c$ control plant users (URLLC users) defined by $\mathcal{N}^c=\{1,2,...,N^c\}$ and a set of $N^e$ eMBB moving users (coexisted with the control plant users) defined by $\mathcal{N}^e=\{1,2,...,N^e\}$. It is noted that URLLC users and control plant users are employed interchangeability from hereon. We consider that URLLC users are fixed and eMBB users move within the range of the BS coverage area. Let dividing the total bandwidth of the network in $J$ sub-carriers forming set $\mathcal{J}=\{1,2,...,J\}$. Also, let $p_{n,j}[k]$ be the base station's transmit power for communicating with downlink user $n$ on sub-carrier $j \in \mathcal{J}$ at discrete time $k$. The variable of $p_{n,j}[k]$ is assumed continues. The overall power transmit of the BS is limited to a maximum value represented by $\overline{P}_{BS}$, which means $\sum_{n=1}^{N}\sum_{j=1}^{J}p_{n,j}[k]\leq\overline{P}_{BS}$. Moreover, the BS' total power usage in the considered ICPS is calculated as \cite{loodaricheh2014energy}:\\
 \begin{equation}
 P^\text{BS}_\text{total}[k]=P_\text{cst}+\epsilon^\text{BS}\sum\nolimits_{n \in \mathcal{N}} \sum\nolimits_{j \in \mathcal{J}}a_{n,j}[k]p_{n,j}[k],
 \label{eqn:poweD}
 \end{equation} 
where $P_{\text{cst}}$ is a constant power used by BS circuit, $\epsilon^\text{{BS}}$ is the amplifier inefficiency constant, $a_{n,j}[k]$ is the sub-carrier allocation variable, which is a binary variable. $a_{n,j}[k]=1$ if sub-carrier $j$ is allocated to downlink user $n$ at discrete time $k$, or else, $a_{n,j}[k]=0$. Also, $\textbf{A}[k]$ and $\textbf{P}[k]$ are power and sub-carrier allocation matrices at discrete time $k$ respectively
($\textbf{A}[k]{\mathop:}=[{a_{n,j}}_{(n\in \mathcal{N},j\in \mathcal{J})}[k]]$ and $\textbf{P}[k]{\mathop:}=[{p_{n,j}}_{(n\in \mathcal{N},j\in \mathcal{J})}[k]]$).\\
The downlink Signal-to-Noise Ratio (SNR) for user $n$ on sub-carrier $j$ is given by \cite{tang2019service}:\\
\begin{equation}
\gamma_{n,j}[k]=\frac{p_{n,j}[k]g_{n,j}[k]}{N_0[k]},
\label{eqn:gammau} 
\end{equation}
where $g_{n,j}[k]$ is the channel gain for each user $n$ on sub-carrier $j$ at discrete time $k$ and $N_0[k]$ denotes the corresponding additive white Gaussian noise power at the receiver of user $n$. 
In accordance with the Shannon's formula, the achievable instantaneous transmission rate for each eMBB user $n \in \mathcal{N}^e$ is computed in bit/s as:\\
\begin{equation}
R^e_n[k]=\sum\nolimits_{j \in \mathcal{J}}w a_{n,j}[k]\log(1+\gamma_{n,j}[k]),
\label{eqn:rateu}
\end{equation} 
where $w$ is the bandwidth of sub-carrier $j$. Moreover, the QoS requirement for each eMBB user $n \in \mathcal{N}^e$ is computed in terms of a minimum transmission rate \cite{tang2019service}. Therefore, the required QoS of eMBB users is represented by:
\begin{equation}	
	R^e_n[k] \geq\overline{R}^e_{n}[k],\forall n\in \mathcal {N}^e,
	\label{eqn:ratecons}
\end{equation}
where $\overline{R}^e_{n}[k]$ is the minimum required QoS of eMBB user $n$ at discrete time $k$. 
The packet size of URLLC users are generally short so the Shannon's formula cannot exactly describe their transmission rate \cite{polyanskiy2010channel,tang2019service}. The achievable transmission rate of URLLC users with the finite blocklength channel coding method is derived in \cite{polyanskiy2010channel} as:
\begin{equation}
\begin{split}
R^c_n[k]\!=&w\!\sum\nolimits_{j \in \mathcal{J}}\!a_{n,j}[k](\log(1+\gamma_{n,j}[k])\\&-\! \sqrt{\!\dfrac{V_{n,j}[k]}{C_{n,j}}}Q^{-1} (\epsilon)\log e),
\label{eqn:rated}
\end{split}
\end{equation}
where $C_{n,j}$ is the number of symbols in each codeword block, $Q^{-1}$ is the inverse of Gaussian Q-function, $\epsilon$ is the
error probability, and $V_{n,j}$ is dispersion of sub-carrier $j$ for user $n \in \mathcal{N} ^c$ given by: 
\begin{equation}
V_{n,j}[k]=1-\dfrac{1}{(1+\gamma_{n,j}[k])^2}.
\label{eqn:V}
\end{equation}
In a single time slot $k$, to satisfy the required QoS of URLLC users, it is necessary to provide the achievable instantaneous data rate condition as below:
\begin{equation}
R^c_n [k]\geq \dfrac{L_c}{T_c[k]},\forall n\in \mathcal {N}^c,
\label{eqn:ratecon_urllc}
\end{equation}
where $L_c$ is the length of actuator's packet size in bits and $T_c[k]$ is the maximum tolerable transmission delay for the packet. We calculate $T_c[k]$ according to the given maximum tolerable end-to-end (e2e) delay between the controller and actuators.  
Let $T^\text{{comp}}_\text{{max}}$ be the maximum queuing and computation delay that is $T^\text{{comp}}[k]\leq T^\text{{comp}}_\text{{max}}$ and the propagation delay is negligible. Thus, we conservatively assume the e2e delay is:
 \begin{equation}
 T_\text{e2e}[k]=T_c[k]+T^\text{{comp}}_\text{{max}}.
 \label{eqn:e2edelay}
 \end{equation} 
Noticeably, we assume the minimum reliability requirement for URLLC users is satisfied through some enabler techniques such as low-rate codes.
\vspace{-5mm}
\subsection{Problem formulation} 
We now formally state the joint design problem of the event-triggered control and the energy-efficient resource allocation of the OFDMA network, as a multi-objective optimization problem. It aims to minimize both the number of updates on the actuators' input and the total downlink power usage, subject to the dynamics and UUB stability of the control plant, the QoS demands of eMBB and URLLC users, power and sub-carrier constraint, and the maximum practicable level of the BS' transmit power. This problem is formulated as:\\
\begin{equation}
\begin{aligned}
\begin{split}
&\underset{\{\beta[k]\},\{u[k]\}}{\text{minimize}} \sum_{i=1}^K\beta[i] \quad \\
&\underset{\{\textbf{A}[k]\},\{\textbf{P}[k]\}}{\text{minimize }} \sum_{i=1}^K P^\text{BS}_\text{total}[i]\\
&\text{subject to}:\\
&C_{1}:x[k+1] = f(x[k], u[k], \omega[k])\\
&\quad \quad y[k] = v(x[k]):\forall k \in \mathcal{K}\\
&C_{2}:u[k] = u[k] \beta[k] + u[k-1](1- \beta[k]):\forall k \in \mathcal{K}\\
&C_{3}:||x[0]||<\zeta \Rightarrow ||x[k]||<\rho: \forall k \geq T(\zeta,\rho)\\
&C_{4}: \beta[k] \in\{0,1\} :\forall k \in \mathcal{K}\\
&C_{5}:R^e_n[k]\geq\overline{R}^e_{n}[k] :\forall k \in \mathcal{K}, \forall n\in \mathcal {N}^e\\
&C_{6}:R^c_n[k] \geq \dfrac{L_c}{T_c[k]}\beta[k] :\forall k\in \mathcal{K}, \forall n\in \mathcal {N}^c\\
&C_{7}:\sum_{n=1}^{N}a_{n,j}[k]\leq 1 :\forall k\in \mathcal{K}, \forall j\in \mathcal {J}\\
&C_{8}:a_{n,j}[k]\in\{0,1\} :\forall n\in \mathcal {N},j\in \mathcal {J},k\in \mathcal{K}\\
&C_{9}:\sum_{n=1}^\mathcal{N}\sum_{j=1}^\mathcal{J}p_{n,j}[k]a_{n,j}[k]\leq\overline{P}_{BS} :\forall k\in \mathcal{K},
\label{eqn:optt}
\end{split}
\end{aligned}
\end{equation}
where constraints $C_{1}$, $C_{2}$, and $C_{3}$ illustrate the plant dynamics, the event-triggered controller function, and UUB stability requirement of the control plant respectively. Constraint $C_{4}$ shows update variable $\beta[k]$ takes binary value. $C_{5}$ and $C_{6}$ represent the required QoS of eMBB and URLLC users respectively.   
Constraints $C_{7}$ and $C_{8}$ are related to the exclusive assignment of the sub-carrier in the OFDMA network. And constraint $C_{9}$ shows the maximum allowable transmit power of the BS. \\
In multi-objective optimization problem (\ref{eqn:optt}), thanks to minimizing the second objective, the transmit power of the control plant's users is reduced. Consequently, the downlink transmission rates are reduced and the transmission delay is increased. Accordingly, to guarantee UUB stability of the control plant  ($C{3}$), the number of updates on the actuators' input is increased in future time steps and the first objective function is increased. Due to the trade-off between these two objective functions, the idea of the Pareto optimality is employed as a solution for problem (\ref{eqn:optt}) \cite{boyd2004convex}. The Pareto optimal solution is defined as follows.
\begin{def2}\label{D2}
	Assuming a multi-objective optimization problem with $f_{i}(k), i \in \{1,2,...,I\}$, as its objective functions and considering all objectives are minimizing functions, a feasible solution, $k^*$, can dominate another one, $k^{**}$, (or $k^*$ is better than $k^{**}$) if:
	\begin{enumerate} 
		\item $f_{i}(k^*)\leq f_{i}(k^{**})$ for all $i\in \{1,2,...,I\}$ and
		\item $f_g(k^*)< f_{g}(k^{**})$ for at least one $g\in \{1,2,...,I\}$.
	\end{enumerate} 
	$k^*$ is named as a Pareto optimal solution when any other solution cannot be found to dominate $k^*$. In other words, $k^*$ is a Pareto optimal solution if and only if it is a feasible solution and there exists no better feasible solution.
\end{def2}   
\vspace{-10pt}
\section{The Proposed approach}
In optimization problem (\ref{eqn:optt}), the dynamics model of the control plant and its interconnection with the network is unknown. To address this problem, we propose a novel model-free HRL approach. Specifically, a Markov Decision Process (MDP) is first constructed associated with problem (\ref{eqn:optt}). Due to the state and action spaces of the MDP are large, we first apply CAN method and decompose problem (\ref{eqn:optt}) into two sub-problems. Then, DAC architecture is used for solving each sub-problem with a hybrid action space.
\vspace{-10pt}
\subsection{RL-related Definition}
The joint design problem can be described by MDP $\mathcal{M}= (\mathcal{S}; \mathcal{A}; \mathcal{R}; \mathcal{P}_0; \mathcal{P}_{ss^{'}})$, where $\mathcal{S}$ is the set of possible states, $\mathcal{A}$ is the set of actions, $\mathcal{R}$ is a reward function ($\mathcal{S} \times \mathcal{A} \rightarrow \mathbb{R}$), $\mathcal{P}_0$ is an initial distribution ($\mathcal{S} \rightarrow [0,1]$), $\mathcal{P}_{ss^{'}}$ is the probability of states transition ($\mathcal{S} \times \mathcal{A}\times \mathcal{S} \rightarrow [0,1]$). The state at time step $k$, $\mathcal{S}[k]$, is defined as:
\begin{equation}
\mathcal{S}[k] = \{s^{c}[k], s^{N}_n[k] \},
\label{eqn:state}
\end{equation}
where $s^{c}[k] = y[k]$, $s^{N}_n[k] \in\{0,1\}: n \in\mathcal{N}$ denotes status of URLLC and eMBB users at environment time step $k$, which $s^{N}_n[k] = 1$ if user $n$ receives its minimum required rate; otherwise $s^{N}_n[k] =0$. We consider the learning agent action at time step $k$, $\mathcal{A}[k]$, as follows: 
\begin{equation}
\mathcal{A}[k] = \{\beta[k], u[k], \textbf{A}[k], \textbf{P}[k]\}.
\label{eqn:action}
\end{equation}
An action is taken, at each time step $k$, on the basis of policy $\pi(\mathcal{A}[k]| \mathcal{S}[k])$, which is a likelihood function of each action for every possible state. By choosing $\mathcal{A}[k]$, the environment state is transmitted from current state $\mathcal{S}[k]$ to $\mathcal{S}[k+1]$ according to the probability of $\mathcal{P}(\mathcal{S}[k+1] | \mathcal{S}[k], \mathcal{A}[k])$ and also a reward of $\mathcal{R}[k+1]$ is gotten ($\mathbb{E}(\mathcal{R}[k+1])= \mathcal{R}(\mathcal{S}[k],\mathcal{A}[k])$). Assuming the transition trajectory as $\tau = (\mathcal{S}[0], \mathcal{A}[0], ..., \mathcal{S}[K])$, the goal of RL is to obtain a policy (${\pi^0}$), which maximize the expected receiving cumulative reward trough the trajectory, which is given by $\mathcal{R}(\tau) = \mathbb{E}(\sum_{j=k}^{K}\gamma^{j-k}\mathcal{R}[k])$ 
where $0 \leq \gamma  < 1$ denotes the discount factor showing the important weight of future rewards. $\mathcal{R}(\tau)$ is the cumulative reward of an episode between the step of $k$ and the terminal step of $K$.
\vspace{-10pt}
\subsection{Applying CAN Method and Decomposing the Problem}
It is obvious that the size of the state and action spaces of the joint design problem may be too large in practical cases. In such a high-dimensional and complex problem, the speed of learning is considerably reduced. Furthermore, the training process generally consumes an unreasonable amount of computation power in the high-dimensional problem. To manage these challenges, CAN method is used as explained in \cite{chang2020cascade}. In CAN method, the learning process of a complicated problem is decomposed into low-dimensional attribute modules, which are linked in cascade series. The state space of every attribute is determined as minimum as possible provided that the space can completely describe the attribute, indicated by $\mathcal{S}=\{\mathcal{S}^0, \mathcal{S}^1, \mathcal{S}^2,...\}$. Also, every attribute enjoys its own reward function ($\mathcal{R}=\{\mathcal{R}^0, \mathcal{R}^1, \mathcal{R}^2,...\}$). 
Moreover, the transition probability distribution in every attribute is indicated by $\mathcal{P}=\{\mathcal{P}^0, \mathcal{P}^1, \mathcal{P}^2,...\}$.
Although it is shown that CAN method makes the training process significantly faster and more simple in \cite{chang2020cascade}, it is not mathematically proven that applying the decoupling method results in an optimal/sub-optimal solution. Here, however, we demonstrate this through the following lemmas in the case of problem (\ref{eqn:optt}).  
\begin{lemma}\label{L1}
The second objective of problem (\ref{eqn:optt}) is decreasing with respect to $\beta[k]$.
\end{lemma}
\begin{IEEEproof}
By decreasing $\beta[k]$, the number of control users that require to communicate  decreases, through which the number of downlink users, $\textit{N}$, is decreasing. Consequently, the total power consumption of the BS will decrease in accordance with equation (\ref{eqn:poweD}).
\end{IEEEproof}
\begin{lemma}\label{L2}
Assuming $\mathcal{H}=\{\beta^*,u,\textbf{A},\textbf{P}\}$ be a subset of feasible solution set of problem (\ref{eqn:optt}), if $\beta^*\in \mathcal{H}$ minimize the first objective function, then $\mathcal{H}$ is a subset of the Pareto solution set of optimization problem (\ref{eqn:optt}).
\end{lemma}
\begin{figure}[t]
	\centerline{\includegraphics[width=0.5\textwidth]{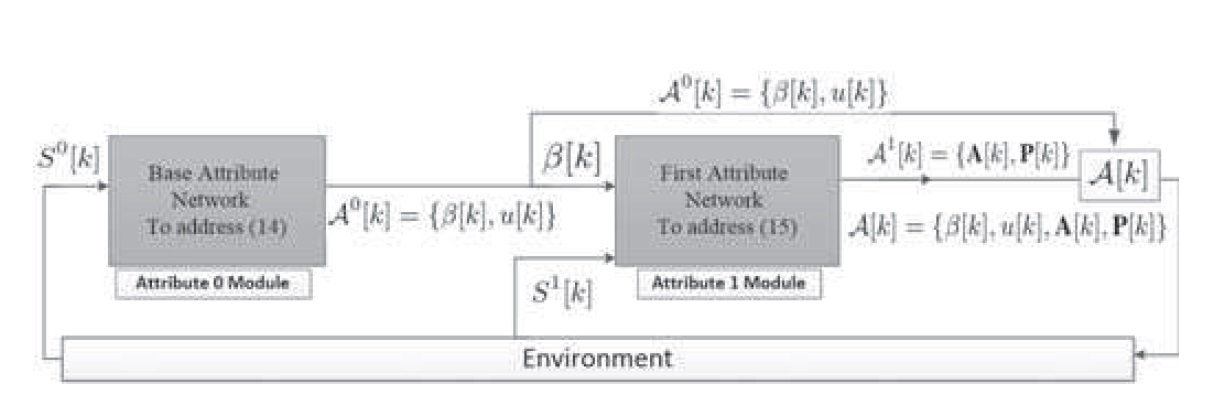}}
	\label{fig:Proposed approach}
	\caption{Proposed approach}
\end{figure} 
\begin{IEEEproof}
Lemma \ref{L2} will be demonstrated by the contradiction. Assuming that there is a feasible solution $k^{**}$, which dominant $k^*\in\mathcal{H}=\{\beta^*,u,\textbf{A},\textbf{P}\}$ ($\beta^*$ minimize the first objective function). But, 
in accordance with Lemma \ref{L1}, the second objective function is decreasing by decreasing $\beta$ and also the first objective is optimized in $\beta^*$. Thus, the conditions presented in Definition 2 are not fulfilled. Accordingly, $k^{**}$ does not dominate $k^*$. As a result, the initial presumption that there is a feasible point, which dominate $k^{*}$ is contradicted. 	
\end{IEEEproof}
Lemma \ref{L2} allows us to decouple optimization problem (\ref{eqn:optt}) into two sub-problems as:
\begin{equation}
\begin{aligned}
&\underset{\{\beta[k]\},\{u[k]\}}{\text{minimize}} \sum_{i=1}^K\beta[i] \\ 
&\text{subject to}:
C_{1}, C_{2}, C_{3}, C_{4},\quad \quad \quad 
\label{eqn:sub1}
\end{aligned}
\end{equation}
\text{and}
\begin{equation}
\begin{aligned}
&\underset{\{\textbf{A}[k]\},\{\textbf{P}[k]\}}{\text{minimize }} \sum_{i=1}^K P^\text{BS}_\text{total}[i]\\
&\text{subject to}: C_{4}, C_{5}, C_{6}, C_{7}, C_{8}, C_{9}.
\label{eqn:sub2}
\end{aligned}
\end{equation}
The architecture of the proposed approach applying CAN method is shown in Fig. 2. The training process of the proposed approach has two parts. In the first part, DRL policy of the base attribute module is trained to address sub-problem (\ref{eqn:sub1}). The base module is fed with $\mathcal{S}^0\subset \mathcal{S}$ and output $\mathcal{A}^0 \subset \mathcal{A}$, considering reward function $\mathcal{R}^0$. Notably, $\mathcal{A}^0$ contains $\beta[k]$ and $u[k]$, which $u[k]$ is a continues variable and $\beta[k]$ is a binary variable. Having decided $\beta[k]$, DRL policy of the first attribute module is trained subsequently, which is accountable to solve sub-problem (\ref{eqn:sub2}). This module is fed with $\mathcal{S}^1 \subset \mathcal{S}$ and output power matrix $\textbf{P}[k]$ along with sub-carrier matrix $\textbf{A}[k]$, considering reward function $\mathcal{R}^1$.\\
The action space of each sub-problem is a hybrid space, and the majority of regular RL-based solutions are not appropriate to solve these hybrid problems \cite{funk2021learning}. Therefore, to address each sub-problem, we propose to use option-critic method, which is reformulated as DAC architecture in \cite{zhang2019dac}, since it is well-suited to deal with hybrid action space \cite{hou2020data, funk2021learning}.
\vspace{-3mm}
\subsection{The Base Attribute Module}
To handle sub-problem (\ref{eqn:sub1}), the state and action spaces of the base module are defined as:
\begin{equation}
\begin{split}
\mathcal{S}^0[k] &= \{s^{c}[k]\},\\
\mathcal{A}^0[k] &= \{\beta[k], u[k]\}.
\label{eqn:state_base}
\end{split}
\end{equation}
The base module is responsible for learning a policy (${\pi^0}$) over $\beta[k]$ and $u[k]$. The policy aim to maximize the expected receiving cumulative reward through transmission trajectory $\tau^0=\{\mathcal{S}^0[0],\beta[0], u[0], \mathcal{S}^0[1],\beta[1], u[1], ..., \mathcal{S}^0[K]\}$. The reward function of the base module is defined as:
	\begin{equation}
	\!\mathcal{R}^{0}(\mathcal{S}^0[k],\mathcal{A}^0[k])\!= \mathcal{R}^{ctrl}[k] -\mu_1\beta[k],
	\label{eqn:reward_base}
	\end{equation}
where the first term ($\mathcal{R}^{ctrl}[k]$) is the control reward and the second term ($-\mu_1\beta[k]$) is to minimize the number of updates on the actuators' input. The control reward is defined to encourage the control plant to reach its specified targets. Also, $\mu_1$ is a hyper-parameter denoting the penalty weight of the number of actuators updates.\\ To guarantee UUB stability of the learning controller with policy ${\pi^0}$, we use a more general definition of UUB stability presented in \cite{han2021reinforcement}. Indeed, in \cite{han2021reinforcement}, the classical definition of UUB stability (Definition 1) is extended for general cases in which the stability constraint functions are not necessarily the norm of the control state ($||x[k]||$). 
Let $\mathcal{C}_{\pi^0}(\mathcal{S}^0[k])\doteq \mathbb{E}_{\mathcal{A}^0[k]\sim{\pi^0}}\mathcal{C}(\mathcal{S}^0[k],\mathcal{A}^0[k])$ be the constraint function under the policy ${\pi^0}$ and $\mathcal{C} (\mathcal{S}^0[k],\mathcal{A}^0[k])$ be a continuous nonnegative constraint function, which is defined to measure how good or bad a state$-$action pair of the base module is. The general definition of UUB stability with respect to $\mathcal{C}_{\pi^0}(.)$ is stated in what follows. 
	\begin{def3}\label{D4}
		A control plant is UUB with respect to $\mathcal{C}_{\pi^0}(.)$, if there are positive constants $b, \rho$ and $\forall \zeta < b$ : $\exists T(\zeta,\rho)$, such that $\mathcal{C}_{\pi^0}(\mathcal{S}^0[k_0])<\zeta \Rightarrow\mathcal{C}_{\pi^0}(\mathcal{S}^0[K])<\rho, \forall k \geq k_0 + T$.	
	\end{def3}  
It is shown that Definition 3 is an inherent feature of the control plant when it is UUB stable. Thus, if the control plant is UUB with respect to $\mathcal{C}_{\pi^0}(.)$, then the closed-loop control is UUB \cite{han2021reinforcement, zhang2021safe}. It is noted that UUB points to the property defined by Definition 3 from hereon.
\begin{theorem1}\label{T1}
	Assuming that the Markov chain induced by policy $\pi^0$ is ergodic, $\Lambda()\doteq\{\mathcal{S}^0[k]\in \mathcal{S}^0 |\mathcal{C}_{\pi^0}(\mathcal{S}^0[k])\geqslant\rho\}$, and $\mathbb{I}_{\Lambda}(\mathcal{S}^0[k])=\left\{
	\begin{array}{rl}
	&1 \quad\quad \qquad \mathcal{S}^0[k]\in\Lambda\\
	&0 \quad \quad\qquad \mathcal{S}^0[k]\notin\Lambda
	\end{array}\right.$, if there are a function $\Gamma(\mathcal{S}^0[k]):\mathcal{S}^0\rightarrow \mathbb{R}^+$ and positive constants $\alpha_1, \alpha_2, \alpha_3$, and $\rho$, such that 
	\begin{equation}
	\alpha_1\mathcal{C}_{\pi^0}(\mathcal{S}^0[k])\leq\Gamma(\mathcal{S}^0[k]) \leq\alpha_2\mathcal{C}_{\pi^0}(\mathcal{S}^0[k]), \forall \mathcal{S}^0[k]\in \mathcal{S}^0,  
	\end{equation}
	and
	\begin{equation}
	\begin{aligned}
	&\mathbb{E}_{\mathcal{S}^0[k]\sim\Omega_{\overline{\rm K}}}(\mathbb{E}_{\mathcal{S}^0[k+1]\sim\mathcal{P}^0}\Gamma(\mathcal{S}^0[k+1])\mathbb{I}_{\Lambda}(\mathcal{S}^0[k+1])-\\
	&\Gamma(\mathcal{S}^0[k])\mathbb{I}_{\Lambda}(\mathcal{S}^0[k]))
	<-\alpha_3 \mathbb{E}_{\mathcal{S}^0[k]\sim\Omega_{\overline{\rm K}}}\mathcal{C}_{\pi^0}(\mathcal{S}^0[k])\mathbb{I}_{\Lambda}(\mathcal{S}^0[k]),
	\end{aligned}
	\label{eqn:stability}	
	\end{equation}
where $\Omega_{\overline{\rm K}}$ shows the average distribution of $\mathcal{S}^0[k]$ over the finite ${\overline{\rm K}}$ time steps, 
		$\Omega_{\overline{\rm K}}(\mathcal{S}^0[k])\doteq \dfrac {1}{\overline{\rm K}} \sum_{k=1}^{{\overline{\rm K}}} \mathcal{P}^0(\mathcal{S}^0[k]|\mathcal{P}_0^0,\pi^0,k)$,
and ${\overline{\rm K}} = max \{k:\mathcal{P}^0(\mathcal{S}^0[k]\in\Lambda|\mathcal{P}_0^0,\pi^0,k)>0\}$, 
then $\pi^0(\mathcal{A}^0[k]| \mathcal{S}^0[k])$ guarantees UUB stability of the control plant with ultimate bound $\rho$. If for any $\epsilon$, there is a $k>\epsilon$, such that  $\mathcal{P}^0(\mathcal{S}^0[k]\in\Lambda|\mathcal{P}_0^0,\pi^0,k)>0$, then ${\overline{\rm K}}= \infty$. 
\end{theorem1}
Similar to \cite{han2021reinforcement}, a fully connected deep neural network is used to construct function $\Gamma_C(\mathcal{S}^0[k],\mathcal{A}^0[k])$, which satisfies $\Gamma(\mathcal{S}^0[k])= \mathbb{E}_{\mathcal{A}^0[k]\sim\pi^0}\Gamma_C(\mathcal{S}^0[k],\mathcal{A}^0[k])$ and the function $\Gamma_C$ is parameterized by $\upsilon$. A ReLU activation function is employed in the output layer of the deep neural network to guarantee positive output. To update $\upsilon$, the
following objective function is minimized:
\begin{equation}
\mathcal{L}(\upsilon)=\mathbb{\tilde{E}}_{\Lambda} (\dfrac{1}{2} (\Gamma_C(\mathcal{S}^0[k],\mathcal{A}^0[k])-\mathcal{C} (\mathcal{S}^0[k],\mathcal{A}^0[k]))^2),\\
\label{eqn:upsilon}
\end{equation}
where $\mathbb{\tilde{E}}_{\Lambda}(.)$ is the average over a mini-batch of samples collected from the sampling distribution $\Omega_{\overline{\rm K}}(s)$.\\
In the following, an approach based on option-critic method, which is reformulated as DAC architecture, is proposed to obtain $\pi^0$. In the obtaining procedure of policy $\pi^0$, we employ Theorem 1 to guarantee UUB stability.\\
\textbf{Option-Critic Method}: 
Option-critic method is an HRL that has three policies: a master policy, an intra$-$option policy, and an option termination function \cite{zhang2019dac, funk2021learning}. The master policy decides which option should be performed. On the basis of this decision, an action is taken through intra$-$option policy until the option is terminated by the termination function. Accordingly, in the context of sub-problem (\ref{eqn:sub1}), the master policy specifies the probability of choosing update variable $\beta[k]$ at each time step $k$ and then control input $u[k]$ is determined by the intra$-$option policy. Furthermore, the termination function is omitted (similar to \cite{hou2020data} and \cite{funk2021learning}) because of the binary type of update variable $\beta[k]$. Indeed, when the master policy chooses one option ($\beta[k] = 1$ or $\beta[k] = 0$), it terminates another option simultaneously. Considering this performing model, we have:
 \begin{equation}
 \begin{split}
\mathcal{P}^0(\mathcal{S}^0[k+1] | \mathcal{S}^0[k], \beta[k]) = \sum_{a} \pi (a=&u[k] | \mathcal{S}^0[k],\beta[k])\times\\ 
\mathcal{P}^0(\mathcal{S}^0[k&+1] | \mathcal{S}^0[k],u[k]),\\
\mathcal{P}^0(\mathcal{S}^0[k+1], \beta[k+1]| \mathcal{S}^0[k], \beta[k]) =  \\\mathcal{P}^0 ( \mathcal{S}^0[k+1] | \mathcal{S}^0[k],\beta[k])\times
 \mathcal{P}^0(\beta [k+1]&| \mathcal{S}^0[k+1], \beta[k]).
 \label{eqn:proab}
 \end{split}
 \end{equation}
In \cite{zhang2019dac}, it is demonstrated that option-critic method can be reformulated as DAC architecture, which contains two augmented MDPs. The MDPs contain the high-level MPD,  $\mathcal{M}^{H_0}$, and the low-level MPD, $\mathcal{M}^{L_0}$, which are employed for choosing the option and the action respectively. Consequently, the high-level MPD of the base module is defined as:
\begin{equation}
\begin{split}
\mathcal{M}^{H_0}\!&\doteq \{\mathcal{S}^{H_0}; \mathcal{A}^{H_0}; \mathcal{R}^{H_0}; \mathcal{P}^{H_0}_0; \mathcal{P}^{H_0}\},\\
\mathcal{S}^{H_0}[k]\!&\doteq \{\beta[k-1], s^{c}[k] \},\\
\mathcal{A}^{H_0}[k]\!&\doteq \{\beta[k] \},\\
\mathcal{R}^{H_0}\!&\doteq \mathcal{R}^{H_0}(\mathcal{S}^{H_0}[k], \mathcal{A}^{H_0}[k])\\ &\doteq \mathcal{R}^0(\mathcal{S}^0[k], \beta[k]),\\
\mathcal{P}^{H_0}_0(\mathcal{S}^{H_0}[0])\!&\doteq \mathcal{P}_0 (\beta[-1], s^{c}[0]),\\
\mathcal{P}^{H_0} (\mathcal{S}^{H_0}[k+1]|\mathcal{S}^{H_0}[k],\mathcal{A}^{H_0}[k])\!&\doteq\!\textbf{1}_{\mathcal{A}^{H_0}=\beta[k]}\mathcal{P}^0 (\mathcal{S}^0[k+1]|\\&\qquad \mathcal{S}^0[k],\!\beta[k]),
\label{eqn:highlevel_C}
\end{split}
\end{equation}
where $\textbf{1}_{(.)}$ is the indicator function. Also, the high-level policy on $\mathcal{M}^{H_0}$ is defined as:
\begin{equation}
\pi^{H_0} (\mathcal{A}^{H_0}[k] | \mathcal{S}^{H_0}[k]) \doteq \mathcal{P}^0 (\beta[k] | \beta[k-1], s^{c}[k]).
\label{eqn:policy_C}
\end{equation}
The low-level MPD and policy of the base module are respectively stated as:
\begin{equation}
\begin{split}
\mathcal{M}^{L_0} &\doteq \{\mathcal{S}^{L_0}; \mathcal{A}^{L_0}; \mathcal{R}^{L_0}; \mathcal{P}^{L_0}_0; \mathcal{P}^{L_0}\},\\
\mathcal{S}^{L_0}[k] & \doteq \{s^{c}[k]\}\times \{\beta[k]\},\\
\mathcal{A}^{L_0}[k] & \doteq \{u[k] \},\\
\mathcal{R}^{L_0} & \doteq \mathcal{R}^{L_0}(\mathcal{S}^{L_0}[k], \mathcal{A}^{L_0}[k]) \\&\doteq \mathcal{R}^0(\mathcal{S}^0[k], \mathcal{A}^0[k]),\\ 
\mathcal{P}^{L_0}_0(\mathcal{S}^{L_0}[0]) & \doteq \mathcal{P}_0 (\mathcal{S}^0[0])\mathcal{P}_0(\beta[0]|\mathcal{S}^0[0]),\\
\mathcal{P}^{L_0} (\mathcal{S}^{L_0}[k+1]|\mathcal{S}^{L_0}[k],\mathcal{A}^{L_0}[k])& \doteq\\
\mathcal{P}^0 ((\beta[k+1], \mathcal{S}^0[k+1])| (\beta&[k], \mathcal{S}^0[k]), \mathcal{A}^{L_0}=u[k]) =\\
 \mathcal{P}^0 (\mathcal{S}^0[k+1]|\mathcal{S}^0[k],u[k]) \times\mathcal{P}^0 &(\beta[k+1] | \mathcal{S}^0[k+1],\beta[k]),
\label{eqn:Lowlevel}
\end{split}
\end{equation}
and
\begin{equation}
\pi^{L_0} (\mathcal{A}^{L_0}[k] | \mathcal{S}^{L_0}[k]) \doteq \mathcal{P}^0 (u[k] | \mathcal{S}^0[k],\beta[k]).
\label{eqn:policy_L}
\end{equation}
Considering trajectories of $\psi^0 = \{\tau^0| \mathcal{P}^0(\tau^0 | \pi^0 ,\mathcal{M}^0)\}$, $\psi^{H_0} = \{\tau^{H_0}| \mathcal{P}^{H_0}(\tau^{H_0} | \pi^{H_0} ,\mathcal{M}^{H_0})\}$ and $\psi^{L_0} = \{\tau^{L_0}| \mathcal{P}^{L_0}(\tau^{L_0} | \pi^{L_0} ,\mathcal{M}^{L_0})\}$, two bijection functions as $\mathcal{J}^{H_0} $ and $\mathcal{J}^{L_0} $ are obtained, which map $\tau^0$ to $\tau^{H_0}$ and $\tau^0$ to $\tau^{L_0}$ respectively. Here, the following lemmas holds, which appear similar to \cite{zhang2019dac}:
\begin{lemma}\label{L3}
Assuming the bijection function $\mathcal{J}^{H_0} $, we have $\mathcal{P}^0(\tau^0 | \pi^0 ,\mathcal{M}^0) = \mathcal{P}^{H_0}(\tau^{H_0} | \pi^{H_0} ,\mathcal{M}^{H_0})$ and $\mathcal{R}^0 (\tau^0)= \mathcal{R}^{H_0} (\tau^{H_0})$.
\end{lemma}
\begin{figure}[h!]
\centerline{\includegraphics[width=0.25\textwidth]{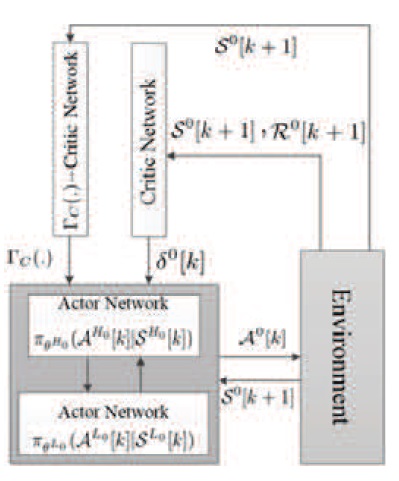}}
\label{fig:actor-critic architecture}
\caption{DAC architecture of the base module}
\end{figure}
\begin{lemma}\label{L4}
	Assuming the bijection function $\mathcal{J}^{L_0} $, we have $\mathcal{P}^0(\tau^0 | \pi^0 ,\mathcal{M}^0) = \mathcal{P}^{L_0}(\tau^{L_0} | \pi^{L_0} ,\mathcal{M}^{L_0})$ and $\mathcal{R}^0 (\tau^0)= \mathcal{R}^{L_0}(\tau^{L_0})$.
\end{lemma}
The proof of above lemmas are provided in Appendices A and B respectively. These lemmas specify that $\{\pi^{H_0} ,\mathcal{M}^{H_0}\}$ and $ \{\pi^{L_0} ,\mathcal{M}^{L_0}$\} can share the same samples with $\{\pi^0 ,\mathcal{M}^0\}$. In the same way of the provided proof, Theorem 2 can be simply driven as follows.
\vspace{-10pt}
\begin{theorem2}
\begin{equation}
\begin{split}
 \Phi^0&\doteq \int \mathcal{R}^0 (\tau^0) \mathcal{P}^0(\tau^0 | \pi^0 ,\mathcal{M}^0)d\tau^0 \\
&= \int \mathcal{R}^{H_0} (\tau^{H_0}) \mathcal{P}^{H_0}(\tau^{H_0} | \pi^{H_0} ,\mathcal{M}^{H_0})d\tau^{H_0}\\
&=  \int \mathcal{R}^{L_0} (\tau^{L_0}) \mathcal{P}^{L_0}(\tau^{L_0} | \pi^{L_0} ,\mathcal{M}^{L_0})d\tau^{L_0}.
\label{eqn:eq-objectives}
\end{split}
\end{equation}
\end{theorem2}
\vspace{-1.5mm} 
\noindent Following Lemma \ref{L3}, Lemma \ref{L4}, and Theorem 2, to handle sub-problem (\ref{eqn:sub1}), the learning agent alternately optimize $\pi^{H_0}$ (decide on option variable $\beta[k]$) and $\pi^{L_0}$ (decide on continues variable $u[k]$). Thereby option-critic method is reformulated to DAC architecture (see Fig. 3). To optimize policies in each augmented MDP ($\mathcal{M}^{H_0}$, $\mathcal{M}^{L_0}$), Proximal Policy Optimization (PPO) method is used similar to \cite{zhang2019dac}.\\ 
In DAC architecture, two parameterized polices $\pi^{H_0}(\mathcal{S}^{H_0}[k], \mathcal{A}^{H_0}[k],\theta^{H_0})$, which is summarized as $\pi_{\theta^{H_0}}(\mathcal{S}^{H_0}[k], \mathcal{A}^{H_0}[k])$ and $\pi^{L_0}(\mathcal{S}^{L_0}[k], \mathcal{A}^{L_0}[k],\theta^{L_0})$, which is summarized as $\pi_{\theta^{L_0}}(\mathcal{S}^{L_0}[k], \mathcal{A}^{L_0}[k])$ are estimated in two actor neural networks. Additionally, the parameterized value function ($V^{0}(\mathcal{S}^{0}[k],\omega^{0})$) and parameterized function $\Gamma_C(\mathcal{S}^0[k],\mathcal{A}^0[k],\upsilon)$ are estimated in two critic neural networks. 
 Therefore, to update two parameters $\theta^{H_0}$ and $\theta^{L_0}$, considering UUB stability constraint, the following objective functions are minimized respectively:
\begin{equation}
\begin{split}
\mathcal{L}_{H_0}(\theta^{H_0}) &= \mathbb{\tilde{E}} [min (\frac{\pi_{\theta^{H_0}}(\mathcal{A}^{H_0}[k] | \mathcal{S}^{H_0}[k])} {\pi_{\theta^{H_0}_{old}}(\mathcal{A}^{H_0}[k] | \mathcal{S}^{H_0}[k])} ,\\& clip (\frac{\pi_{\theta^{H_0}}(\mathcal{A}^{H_0}[k] | \mathcal{S}^{H_0}[k])} {\pi_{\theta^{H_0}_{old}}(\mathcal{A}^{H_0}[k] | \mathcal{S}^{H_0}[k])}, 1-\epsilon, 1+\epsilon)) \widehat{A}_0[k]]\\& + \lambda \mathbb{\tilde{E}}_{\Lambda}[\Gamma_C(\mathcal{S}^0[k+1],\mathcal{A}^0[k+1])\mathbb{I}_{\Lambda}(\mathcal{S}^0[k+1])\\&-(\Gamma_C(\mathcal{S}^0[k],\mathcal{A}^0[k])-\alpha_3\mathcal{C}_{\pi^0}(\mathcal{S}^0[k]))\mathbb{I}_{\Lambda}(\mathcal{S}^0[k])],
\label{eqn:eq-objective func1}
\end{split}
\end{equation}
and
\vspace{-8pt}
\begin{equation}
\begin{split}
\mathcal{L}_{L_0}(\theta^{L_0}) =& \mathbb{\tilde{E}} [min (\frac{\pi_{\theta^{L_0}}(\mathcal{A}^{L_0}[k] | \mathcal{S}^{L_0}[k])} {\pi_{\theta^{L_0}_{old}}(\mathcal{A}^{L_0}[k] | \mathcal{S}^{L_0}[k])} ,\\& clip (\frac{\pi_{\theta^{L_0}}(\mathcal{A}^{L_0}[k] | \mathcal{S}^{L_0}[k])} {\pi_{\theta^{L_0}_{old}}(\mathcal{A}^{L_0}[k] | \mathcal{S}^{L_0}[k])}, 1-\epsilon, 1+\epsilon)) \widehat{A}_0[k]]\\&+\lambda \mathbb{\tilde{E}}_{\Lambda}[\Gamma_C(\mathcal{S}^0[k+1],\mathcal{A}^0[k+1])\mathbb{I}_{\Lambda}(\mathcal{S}^0[k+1])\\&-(\Gamma_C(\mathcal{S}^0[k],\mathcal{A}^0[k])-\alpha_3\mathcal{C}_{\pi^0}(\mathcal{S}^0[k]))\mathbb{I}_{\Lambda}(\mathcal{S}^0[k])],
\label{eqn:eq-objective func2}
\end{split}
\end{equation}
where $\mathbb{\tilde{E}}(.)$ is the average over a mini-batch of samples (the size of the mini-batch is $\mathcal{Y}^0$), function $clip(.)$ constrains the ratio of $\frac{\pi_{\theta^{L_0}}(\mathcal{A}^{L_0}[k] | \mathcal{S}^{L_0}[k])} {\pi_{\theta^{L_0}_{old}}(\mathcal{A}^{L_0}[k] | \mathcal{S}^{L_0}[k])}$ between the interval of $(1-\epsilon, 1+\epsilon)$, and $\epsilon$ is a hyper-parameter. $\lambda$ is a positive Lagrangian multiplier, which is adjusted via gradient ascent to maximize the following objective function \cite{han2021reinforcement}:
\begin{equation}
\begin{split}
\mathcal{L}(\lambda) = &\lambda \mathbb{\tilde{E}}_{\Lambda}[\Gamma_C(\mathcal{S}^0[k+1],\mathcal{A}^0[k+1])\mathbb{I}_{\Lambda}(\mathcal{S}^0[k+1])\\&-(\Gamma_C(\mathcal{S}^0[k],\mathcal{A}^0[k])-\alpha_3\mathcal{C}_{\pi^0}(\mathcal{S}^0[k]))\mathbb{I}_{\Lambda}(\mathcal{S}^0[k])].
\label{eqn:lambda}
\end{split}
\end{equation}In equations (\ref{eqn:eq-objective func1}) and (\ref{eqn:eq-objective func2}), $\widehat{A}_0[k]$ is the advantage function at time step $k$ and is estimated via the Generalized Advantage Estimation (GAE) as:
\begin{equation}
\begin{split}
\widehat{A}_0[k] = & \delta^0 [k] + (\gamma \xi)\delta^0 [k+1] (\gamma \xi)\delta^0 [k+1]+...\\&+(\gamma \xi)^{(\mathcal{Y}^0-k+1)}\delta^0 [\mathcal{Y}^0-1],
\label{eqn:advantag_0}
\end{split}
\end{equation}
where $\xi \in [0,1]$ is the GAE parameter and $\delta^0[k]$ is the temporal difference(TD) error, given by $\delta^0[k] = \mathcal{R}^0[k+1] +\gamma V^0_{\omega^0} (\mathcal{S}^0[k+1]) - V^0_{\omega^0} (\mathcal{S}^0[k])$. $\omega^0$ is updated by an Stochastic Gradient Descent (SGD) algorithm as:
\begin{equation}
\omega^0 = \omega^0 - \zeta_{\omega^0} \nabla L^V(\omega^0),
\label{eqn:updat_omaga}
\end{equation}
where $\zeta_{\omega^0}$ is the learning rate and $L^V(\omega^0)$ is the objective function calculated as:\\
\begin{equation}
L^V(\omega^0) = \mathbb{\tilde{E}}[ \vert \widehat{V}^{\text{target}}_{\omega^0}(\mathcal{S}^0[k] )- V^0_{\omega^0}(\mathcal{S}^0[k] )\vert ],
\label{eqn:objective_omaga}
\end{equation}
where $\widehat{V}^{\text{target}}_{\omega^0}(\mathcal{S}^0[k] ) = \mathcal{R}^0[k+1] + \gamma V^0_{\omega^0}(\mathcal{S}^0[k+1])$ is the target value of time-difference error.\\
In summary, at time step $k$, each actor network selects its action according to its current state using its policy. This leads to the state transition to $\mathcal{S}^{0}[k+1]$ and a new reward value, which is estimated by the critic network via value function $V^{0}(\mathcal{S}^{0}[k],\omega^{0})$. Afterward, the TD error is calculated, which is the critic network feedback to optimize $\pi^{H_0}$ and $\pi^{L_0}$ using optimizing problems (\ref{eqn:eq-objective func1}) and (\ref{eqn:eq-objective func2}). In addition, the selected actions and the state transition ($\mathcal{S}^{0}[k+1]$ and $\mathcal{A}^0[k+1]$) lead to updating $\upsilon$ and $\lambda$ through (\ref{eqn:upsilon}) and (\ref{eqn:lambda}). Then, the updated $\Gamma_C(.)$ is sent to actor networks as feedback of $\Gamma_C(.)-$Critic Network to optimize $\pi^{H_0}$ and $\pi^{L_0}$.
\vspace{-10pt}
\subsection{The First Attribute Module}
Having decided update variable $\beta[k]$, DRL policy of the first attribute module is trained to address sub-problem (\ref{eqn:sub2}). The state and action spaces of this module are given by:
\begin{equation}
\begin{split}
\mathcal{S}^1[k] &= \{s^{N}_n[k] \},\\
\mathcal{A}^1[k] &= \{\textbf{A}[k], \textbf{P}[k]\}.
\label{eqn:state_first}
\end{split}
\end{equation} 
Considering the objective and the constraints of sub-problem (\ref{eqn:sub2}), the reward function of the first module is calculated as:
\begin{equation}
\begin{split}
\mathcal{R}^{1}[k]  = \left\{
\begin{array}{rl}
&-\mu_2 \quad\qquad \overline{P}_\text{BS}\leq\sum_{n=1}^\mathcal{N}\sum_{j=1}^\mathcal{J}p_{n,j}[k]a_{n,j}[k] \\
&-\mathcal{P}^\text{BS}_\text{total}+ \mu_3 \sum_{n=1}^{N}s^{N}_n[k] \quad \quad\qquad o/w,
\end{array}\right.
\label{eqn:reward_first}
\end{split}
\end{equation}
where $\mu_2$ is a hyper-parameter denoted the penalty weight on 
crossing the limitation of the BS' power consumption. Also, $\mu_3$ is a hyper-parameter indicated the weight on the number of users received their required rate.\\
\textbf{Option-Critic Method}: The action space of this module is also hybrid. Accordingly, to address sub-problem (\ref{eqn:sub2}), option-critic method reformulated as DAC architecture is employed too. Assuming sub-carrier allocation matrix $\textbf{A}[k]$ is the option variable, we have:
\begin{equation}
\begin{split}
\mathcal{P}^1(\mathcal{S}^1[k+1] | \mathcal{S}^1[k], \textbf{A}[k]) = \sum_{a} \pi (a=&\textbf{P}[k] | \mathcal{S}^1[k],\textbf{A}[k])\times\\ 
\mathcal{P}^1(\mathcal{S}^1[k&+1] | \mathcal{S}^1[k],\textbf{P}[k]),\\
\mathcal{P}^1(\mathcal{S}^1[k+1], \textbf{A}[k+1]| \mathcal{S}^1[k], \textbf{A}[k]) =  \\\mathcal{P}^1 ( \mathcal{S}^1[k+1] | \mathcal{S}^1[k],\textbf{A}[k])\times
\mathcal{P}^1(\textbf{A}[k+1] &| \mathcal{S}^1[k+1], \textbf{A}[k]).
\label{eqn:proab_first}
\end{split}
\end{equation}
Option-critic method can be reformulated as two augmented MDPs. The high-level MPD ($\mathcal{M}^{H_1}$) is used for the sub-carrier assignment ($\textbf{A}[k]$) and the low-level MPD ($\mathcal{M}^{L_1}$) is used for the power allocation ($\textbf{P}[k]$). The high-level MPD of the first module is given as:
\vspace{-6pt} 
\begin{equation}
\begin{split}
\mathcal{M}^{H_1}\doteq &\{\mathcal{S}^{H_1};\mathcal{A}^{H_1}; \mathcal{R}^{H_1};\\ &\quad \mathcal{P}_0^{H_1};\mathcal{P}^{H_1}\},\\
\mathcal{S}^{H_1}[k]\doteq &\{\textbf{A}[k-1], s^{N}_n[k] \},\\
\mathcal{A}^{H_1}[k]\doteq &\{\textbf{A}[k] \},\\
\mathcal{R}^{H_1}\doteq &\mathcal{R}^{H_1}(\mathcal{S}^{H_1}, \mathcal{A}^{H_1})\\\doteq &\mathcal{R}^1(\mathcal{S}^1[k],  \textbf{A}[k]),\\
\mathcal{P}_0^{H_1}(\mathcal{S}^{H_1}[0])\doteq &\mathcal{P}_0^1 ((\textbf{A}[-1], s^{N}_n[0])),\\
\mathcal{P}^{H_1} (\mathcal{S}^{H_1}[k+1]|\mathcal{S}^{H_1}[k],\mathcal{A}^{H_1}[k] )\doteq&\textbf{1}_{\mathcal{A}^{H_1}[k]=\textbf{A}[k]}\mathcal{P}^1 ( \mathcal{S}^1[k+1]|\\& \mathcal{S}^1[k],\textbf{A}[k]).
\label{eqn:highlevel_Cfirst}
\end{split}
\end{equation}
\vspace{-2pt} 
And the high-level policy of the first module on $\mathcal{M}^{H_1}$ is defined as:
\begin{equation}
\pi^{H_1} (\mathcal{A}^{H_1}[k] | \mathcal{S}^{H_1}[k]) \doteq \mathcal{P}^1 (\textbf{A}[k] | \textbf{A}[k-1], s^{N}_n[k]).
\label{eqn:policy_C_first}
\end{equation}
The low-level MPD and policy of the first module are respectively defined as:
\begin{equation}
\begin{split}
\mathcal{M}^{L_1} \doteq& \{\mathcal{S}^{L_1}; \mathcal{A}^{L_1}; \mathcal{R}^{L_1}; \mathcal{P}^{L_1}_0;\\ & \quad \mathcal{P}^{L_1}\},\\
\mathcal{S}^{L_1}[k] \doteq& \{s^{N}_n[k]\}\times \{\textbf{A}[k]\},\\
\mathcal{A}^{L_1}[k]  \doteq& \{\textbf{P}[k] \},\\
\mathcal{R}^{L_1} \doteq& \mathcal{R}^{L_1}(\mathcal{S}^{L_1}, \mathcal{A}^{L_1})\\=& \mathcal{R}^{L_1}((\mathcal{S}^{L_1},\textbf{A}[k]),\textbf{P}[k]) \\\doteq& \mathcal{R}^1(\mathcal{S}^1[k], \mathcal{A}^1[k]),\\ 
\mathcal{P}^{L_1}_0(\mathcal{S}^{L_1}[0]) \doteq& \mathcal{P}^1_0 (\mathcal{S}^1[0])\mathcal{P}^1_0(\textbf{A}[0]|\mathcal{S}^1[0]),\\
\mathcal{P}^{L_1} (\mathcal{S}^{L_1}[k+1]|\mathcal{S}^{L_1}[k],\mathcal{A}^{L_1}[k]) \doteq&\\
\mathcal{P}^1 ((\textbf{A}[k+1], \mathcal{S}^1[k+1])| (\textbf{A}[k]&, \mathcal{S}^1[k]), \mathcal{A}^{L_1}=\textbf{P}[k]) \\
= \mathcal{P}^1 (\mathcal{S}^1[k+1]|\mathcal{S}^1[k],\textbf{P}[k]) \times\mathcal{P}^1 &(\textbf{A}[k+1] | \mathcal{S}^1[k+1],\textbf{P}[k]),
\label{eqn:Lowlevel_first}
\end{split}
\end{equation}
and
\begin{equation}
\pi^{L_1} (\mathcal{A}^{L_1}[k] | \mathcal{S}^{L_1}[k]) \doteq \mathcal{P}^1 (\textbf{P}[k] | \mathcal{S}^1[k], \textbf{A}[k]).
\label{eqn:policy_L_first}
\end{equation}
Considering trajectories of $\psi^1 = \{\tau^1| \mathcal{P}^1(\tau^1 | \pi^1 ,\mathcal{M}^1)\}$, $\psi^{H_1} = \{\tau^{H_1}| \mathcal{P}^{H_1}(\tau^{H_1} | \pi^{H_1} ,\mathcal{M}^{H_1})\}$ and $\psi^{L_1} = \{\tau^{L_1}| \mathcal{P}^{L_1}(\tau^{L_1} | \pi^{L_1} ,\mathcal{M}^{L_1})\}$, two bijection functions as $\mathcal{J}^{H_1} $ and $\mathcal{J}^{L_1} $ can be found, which map $\tau^1$ to $\tau^{H_1}$ and $\tau^1$ to $\tau^{L_1}$ respectively. Similar lemmas to Lemma \ref{L3} and Lemma \ref{L4} can be stated, which indicate that $\{\pi^{H_1} ,\mathcal{M}^{H_1}\}$ and $ \{\pi^{L_1} ,\mathcal{M}^{L_1}$\} can share the same samples with $\{\pi^1 ,\mathcal{M}^1\}$. Similarly, Theorem 3 can be driven as:
\begin{theorem3}
\begin{equation}
\begin{split}
\Phi^1&\doteq \int \mathcal{R}^1 (\tau^1) \mathcal{P}^1(\tau^1 | \pi^1 ,\mathcal{M}^1)d\tau^1 \\
&= \int \mathcal{R}^{H_1} (\tau^{H_1}) \mathcal{P}^{H_1}(\tau^{H_1} | \pi^{H_1} ,\mathcal{M}^{H_1})d\tau^{H_1}\\
&=  \int \mathcal{R}^{L_1} (\tau^{L_1}) \mathcal{P}^{L_1}(\tau^{L_1}| \pi^{L_1} ,\mathcal{M}^{L_1})d\tau^{L_1}.
\label{eqn:eq-objectives_first}
\end{split}
\end{equation}
\end{theorem3} 
Therefore, to solve sub-problem (\ref{eqn:sub2}), the learning agent alternatively optimize $\pi^{H_1}$ (decide on sub-carrier assignment $\textbf{A}[k]$ in $\mathcal{M}^{H_1}$) and $\pi^{L_1}$ (decide on power allocation $\textbf{P}[k]$ in $\mathcal{M}^{L_1}$). To optimize the policies in each augmented MDP of the first module ($\pi^{H_1}$ and $\pi^{L_1}$), PPO method is employed analogous to the previous sub-section.\\
\indent The proposed approach to address problem (\ref{eqn:optt}) is called Cascade Stable Double Actor-Critic (CSDAC). In summary, Algorithm 1 provides the pseudo-code of CSDAC approach.
\begin{algorithm}\label{nfv-alg}
	\BlankLine		
	\SetKwFunction{Range}{range}
	\SetKw{KwTo}{in}\SetKwFor{For}{for}{\string:}{}%
	\SetKwIF{If}{ElseIf}{Else}{if}{:}{elif}{else:}{}%
	
	\SetAlgoNoEnd
	
	\SetAlgoNoLine%
	\textbf{Input:} The mini-batch sizes of $\mathcal{Y}^{\Lambda}$, $\mathcal{Y}^0$ and $\mathcal{Y}^1$.\\
	Set iteration index $i \leftarrow  0 $.\\
	\textbf{Repeat}\\
	\Indp
	for each time step do:\\
	\Indp
	Observe state $\mathcal{S}^0[k]$.\\
	Execute action $\mathcal{A}^{H_0}[k]$ and $\mathcal{A}^{L_0}[k]$ and get reward $ \mathcal{R}^0[k+1]$ (according to (\ref{eqn:reward_base})) and transmit to $\mathcal{S}^0[k+1]$. \\
	Store transition $(\mathcal{S}^0[k], \mathcal{A}^{0}[k], \mathcal{R}^0[k+1], \mathcal{S}^0[k+1])$ in $\mathcal{D}^0$.\\
	Record the largest instant $\overline{\rm K}$ at which $s\in\Lambda$ and Store transition $(\mathcal{S}^0[k], \mathcal{A}^{H_0}[k], \mathcal{A}^{L_0}[k], \mathcal{C}[k], \mathcal{R}^0[k+1], \mathcal{S}^0[k+1]): k<\overline{\rm K}$ in $\mathcal{D}^{\Lambda} $.\\
	Sample mini-batches of transitions from $\mathcal{D}^0$ and $\mathcal{D}^{\Lambda}$ and for each update step do:\\
	\Indp
	\Indp
	Optimize $\pi_{\theta^{H_0}}$ (update parameter $\theta^{H_0}$ through the minimization of (\ref{eqn:eq-objective func1})).\\
	Optimize $\pi_{\theta^{L_0}}$ (update parameter $\theta^{L_0}$ through the minimization of (\ref{eqn:eq-objective func2})).\\
	Optimize $\Gamma_C(.)$ is (update parameter $\upsilon$ through the minimization of (\ref{eqn:upsilon})).\\
	Update $\omega^0$ (according to (\ref{eqn:updat_omaga})) and Calculate the TD error $\delta^0[k]$.\\
	Optimize $\mathcal{L}(\lambda)$ (update parameter $\lambda$ through the minimization of (\ref{eqn:lambda})).\\
	\Indm
	\Indm
	Observe $\beta[k]$ and $\mathcal{S}^1[k]$.\\
	Execute action $\mathcal{A}^{H_1}[k]$ and $\mathcal{A}^{L_1}[k]$ and get reward  $\mathcal{R}^1[k+1]$ (according to (\ref{eqn:reward_first})) and transmit to $\mathcal{S}^1[k+1]$.\\
	Store transition $(\mathcal{S}^1[k], \mathcal{A}^{H_1}[k], \mathcal{A}^{L_1}[k], \mathcal{R}^1[k+1], \mathcal{S}^1[k+1])$ in $\mathcal{D}^1$.\\
	Sample mini-batches of transitions from $\mathcal{D}^1$ and for each update step do:\\
	\Indp
	\Indp
	Optimize $\pi^{H_1}$ (similar to $\pi_{\theta^{H_0}}$). \\
	Optimize  $\pi^{L_1}$(similar to $\pi_{\theta^{L_0}}$).\\
	Update $\omega^1$ (similar to (\ref{eqn:updat_omaga})) and Calculate TD error $\delta^1[k]$.\\
	\Indm
	\Indm
	\Indm
	end for.\\
	$i \leftarrow i + 1$.\\
	\Indm
	\textbf{until}(\ref{eqn:stability}) is satisfied and $i$ exceeds a designed threshold.
	\caption{Pseudocode of Cascade Stable Double Actor-Critic (CSDAC) }	
\end{algorithm}
\section{Validation Results} 
In this section, we evaluate the general applicability and efficacy of our proposed approach (CSDAC) through various simulation results. First, we show the general applicability of CSDAC through a rather simple simulated ICPS including OpenAI Gym Cart-Pole environment \cite{gym} communicating over a single cell OFDMA. This low-dimensional environment allows us to compare the performance of CSDAC with a disjoint design of control and communication methods based on a classical ETC method. Second, the simulated ICPS consists of PyBullet-Gym Ant environment \cite{pybullet} communicating over a single cell OFDMA. We show the capability of CSDAC in this challenging high-dimensional environment. For both simulated ICPSs, we consider a single cell OFDMA as their network sub-systems with the same specifications. In the downlink model of each cellular network, eMBB and URLLC users are considered randomly dispersed in a square cell. The channel gain for each user $n$ on sub-carrier $j$ at time step $k$ is calculated as $g_{n,j}[k] = hd_n^{-3}[k]$, where $d_n[k]$ is the distance between user $n$ and the BS as well as $h = 0.09$ is the loss factor. The distance is fixed for URLLC users; however, it varies for eMBB users. Moreover, we assume that eMBB users move only within the BS coverage area during the simulation time. The other parameters of the simulated networks are given in Table II.
\begin{table}\begin{center}
		\caption{Values of Simulation Parameters}
		\begin{tabular}{||c  c ||} 
			\hline
			Parameter & Value \\ [0.7ex] 
			\hline\hline
			Number of eMBB users & 8\\
			\hline
			Bandwidth of sub-carrier ($\omega$) & 180 KHz \\ 
			\hline
			Maximum transmission power of the BS ($\overline{P}_{BS}$) & 44 dBm\\
			\hline
			Constant power of the BS ($P_{cst}$) & 20 dBm  \\
			\hline
			Noise power ($N_0[k]$)  & -62 dBm \\
			\hline
			Distance between a user and the BS ($d_n[k]$) & 10-50 m \\
			\hline
			Minimum data-rate for an RC user ($\overline{R}^e_{n}[k]$) & 100 bit/s \\
			\hline
			Packet length of control users ($L_c$)  & 70 bit \\
			\hline
			Number of sub-carriers ($J$)  & 8\\ [1ex] 
			\hline
		\end{tabular}
		\label{tab: Table11}
\end{center}\end{table}
\vspace{-15pt} 
\subsection{Cart-Pole Environment}
\vspace{-3pt} 
Cart-Pole is a classical control problem, which includes a pole placed on a cart. This system is a two-degree-of-freedom system containing the linear movement of the cart through the X axis ($x_d$) and the rotational movement of the pole on the X-Y axes ($x_r$). The goal of the control problem is to keep the pole upright via moving the cart left/right. Each episode of the Gym simulation ends when the rotational movement of the pole is more than 0.261 radians or the linear movement of the cart is more than 2.4 units from the origin.\\  
As a proof of concept, we apply CSDAC, described in Sec.III, to jointly design the event-triggered control and the resource allocation in the simulated ICPS including Cart-Pole environment communicating over the single cell OFDMA. Moreover, the efficacy of CSDAC is compared to a disjoint and model-based method, which is a combination of the suggested methods in  \cite{luo2019periodic} and \cite{kai2019joint}. We consider the control reward ($\mathcal{R}^{ctrl}$) is identical to the default reward provided by the Gym environment. The constraint function is calculated by $\mathcal{C}= max(||x_d[k]||-1.1,0)$. To estimate each hyper-parameter in equations (\ref{eqn:reward_base}) and (\ref{eqn:reward_first}), 25 values between 0.01 and 100 are examined. The efficiency of each hyper-parameter is analyzed through 5 randomized training processes using different random inputs. We evaluate the performance of CSDAC by carrying out 50 randomized test episodes and each episode is terminated after 300 discrete time steps. Fig. 4 and Fig. 5 depict the results from one test episode. Fig. 4 illustrates the control state responses obtained with the learning event-triggered controller. Having 46 updates on control input for 25s, the state variables of Cart-Pole system remain stable within the range.\\
To compare CSDAC to a disjoint and model-based method, we use Matlab/Simulink applying a linear model of Cart-Pole as \cite{shi2009output}:\\
\begin{equation}
\begin{split}
\begin{bmatrix}
\dot{x_d}(t)\\
\ddot {x}_d(t)\\
\dot {x_r}(t)\\
\ddot {x}_r(t)
\end{bmatrix}
\!=& \!\begin{bmatrix}
1 & 0.1 & -0.0166 & -0.0005 \\
0 & 1 & -0.3374 & -0.0166\\ 
0 & 0 & 1.0996 & 0.1033 \\
0 & 0 & 2.0247 & 1.0996 
\end{bmatrix}\!\begin{bmatrix}
x_d(t)\\
\dot {x}_d(t)\\
x_r(t)\\
\dot{x}_r(t) 
\end{bmatrix}\!+\!\begin{bmatrix}
0.0045\\
0.0896\\ 
-0.0068\\
-0.1377\\
\end{bmatrix}\!u,\\
\begin{bmatrix}
x_d(t)\\
x_r(t)
\end{bmatrix} =& \begin{bmatrix}
1 & 0 & 0 & 0 \\
0 & 0 & 1 & 0 \\ 
\end{bmatrix} 
(\begin{bmatrix}
x_d(t),
\dot{x}_d(t),
x_r(t),
\dot{x}_r(t) 
\end{bmatrix})^T,
\label{eqn:pl2}
\end{split}
\end{equation}
where $x_d(t)$ is the horizontal displacement of the cart from origin, $x_r(t)$ is the rotational movement of the pole, $u(t)$ denotes the enforced control input to the cart. We employ the event-triggered method proposed in \cite{luo2019periodic} with the linear quadratic regulation method to guarantee the closed-loop stability and an upper bound of performance, which is defined in \cite{luo2019periodic}. Moreover, the optimal resource allocation is carried out using MATLAB/CVX similar to \cite{kai2019joint}. Indeed, we start with the sub-carrier assignment for a given power matrix. Next, the power matrix is allocated while using the previous step sub-carrier assignment matrix. These two steps are done iteratively up to reach a convergence criteria.\\ 
The simulation results, illustrated in Fig. 4 and Fig. 6, show both the model-based method (in accordance with \cite{luo2019periodic, kai2019joint}) and CSDAC are properly controlled the ICPS. However, the total number of updates on the actuator's input, under CSDAC algorithm and the model-based method, are respectively 46 and 102. 
In other words, the number of updates on the actuator's input is reduced by around 55\% when CSDAC is employed. Also, the downlink power consumption in CSDAC algorithm and the model-based method are compared in Fig. 8, where a roughly 64\% downlink power saving is observed.  
\begin{figure}[h!]
	\centerline{\includegraphics[width=0.5\textwidth]{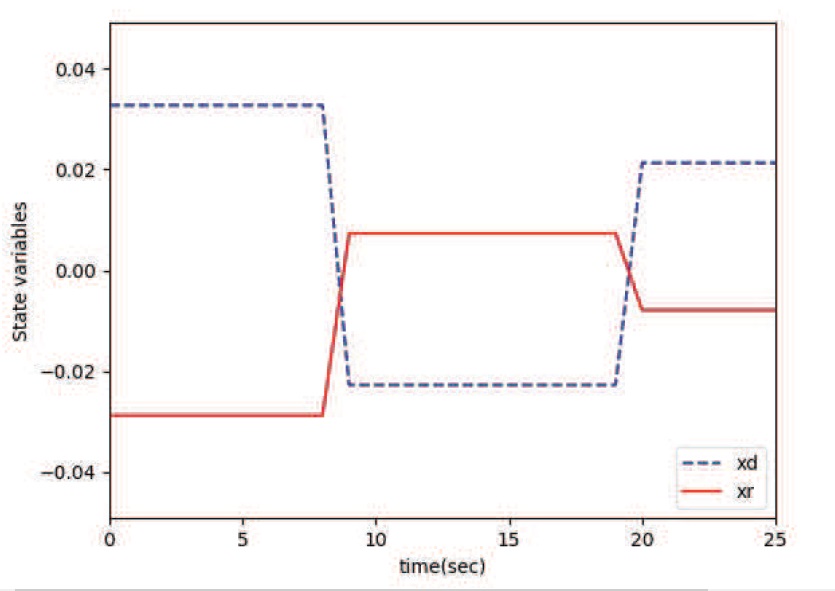}}
	\label{fig:learnig_cartpol}
	\caption{State variables' response obtained with CSDAC}
\end{figure}
\begin{figure}[h!]
	\centerline{\includegraphics[width=0.5\textwidth]{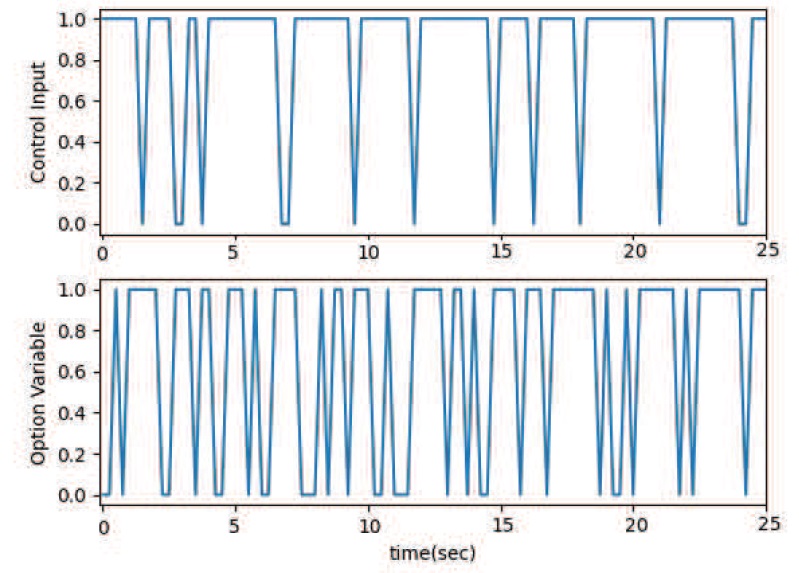}}
	\label{fig:learnig_cartpolg}
	\caption{Variation of control input and option variables in CSDAC}
\end{figure}
\begin{figure}[h!]
	\centerline{\includegraphics[width=0.5\textwidth]{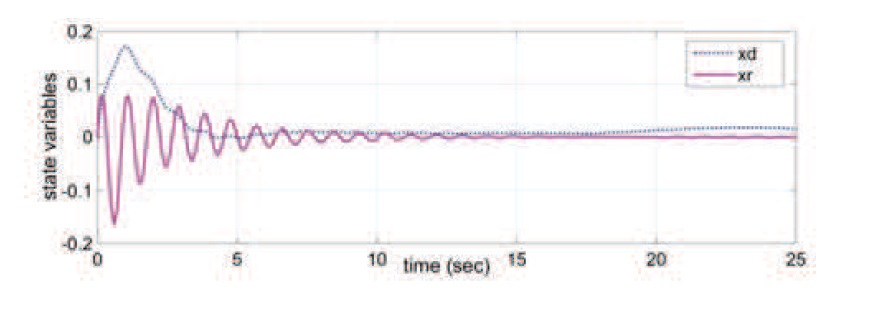}}
	\label{fig:resp model-based event-triggered}
	\caption{State variables' response obtained with the model-based method \cite{luo2019periodic, kai2019joint}}
\end{figure} 
\begin{figure}[h!]
	\centerline{\includegraphics[width=0.5\textwidth]{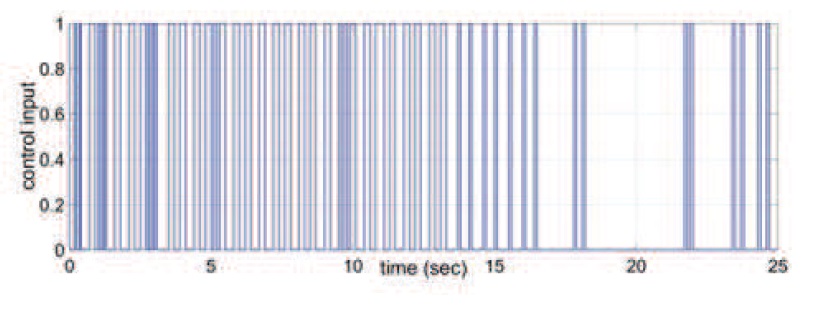}}
	\label{fig:control model-based event-triggered}
	\caption{Variation of the control input in the model-based method \cite{luo2019periodic, kai2019joint}}
\end{figure} 
\begin{figure}[h!]
 	\centerline{\includegraphics[width=0.5\textwidth]{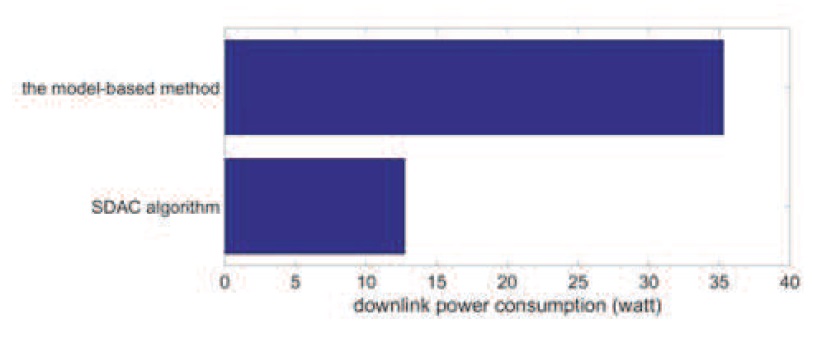}}
 	\label{fig:power_compare}
 	\caption{Compare downlink power consumption in CSDAC with the model-base method (in accordance with \cite{luo2019periodic, kai2019joint})}
 \end{figure}
 \begin{figure}[h!]
	\centering
	\centerline{\includegraphics[width=0.25\textwidth]{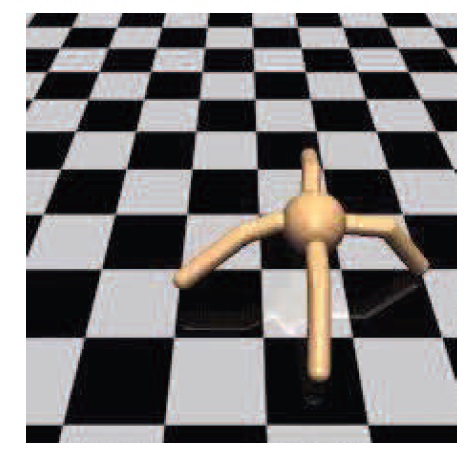}}
	\label{fig:ant}
	\caption{Ant environment screen-shot}
\end{figure}
\subsection{Ant Environment}
We now focus on PyBullet-Gym Ant environment. In such nonlinear and high-dimensional environments, employing known ETC methods is usually unsuccessful as the settings are too complicated \cite{funk2021learning}. 
Ant is a three-dimensional quadruped robot (see Fig. 9), which is rewarded to run forward as fast as possible while a safety constraint on its speed ($v<2.3$) needs to be ensured. We apply CSDAC algorithm in the simulated ICPS including Ant environment communicating over the single cell OFDMA. Moreover, the capability of CSDAC is compared to the soft actor$-$critic (SAC) algorithm \cite{haarnoja2018soft}. We assume the control reward ($\mathcal{R}^{ctrl}$) is the same with
the default reward in the PyBullet-Gym. Also, the safety
constraint on forwarding speed is $\mathcal{C}= max(||v[k]||-2.3,0)$. To estimate each hyper-parameter in equations (\ref{eqn:reward_base}) and (\ref{eqn:reward_first}), we follow the same procedure described in the previous sub-section. Also, each episode is terminated after 300 discrete time steps.\\
As illustrated in Fig. 10, CSDAC performs stably with respect to the safety constraint. On the other hand, SAC fails to find UUB stable policy concerning the safety constraint. The simulation results shown in Fig. 10, Fig. 11, and Fig. 12, demonstrate that the number of constraint violations is approximately zero in CSDAC policy while maintaining reasonable rewards of the base and first modules. In terms of downlink power consumption saving and users' QoS satisfying, as shown in Fig. 12, CSDAC performs much better than SAC and returns significantly more reward in the first module.
\begin{figure}[h!]
	\centering
	\centerline{\includegraphics[width=0.5\textwidth]{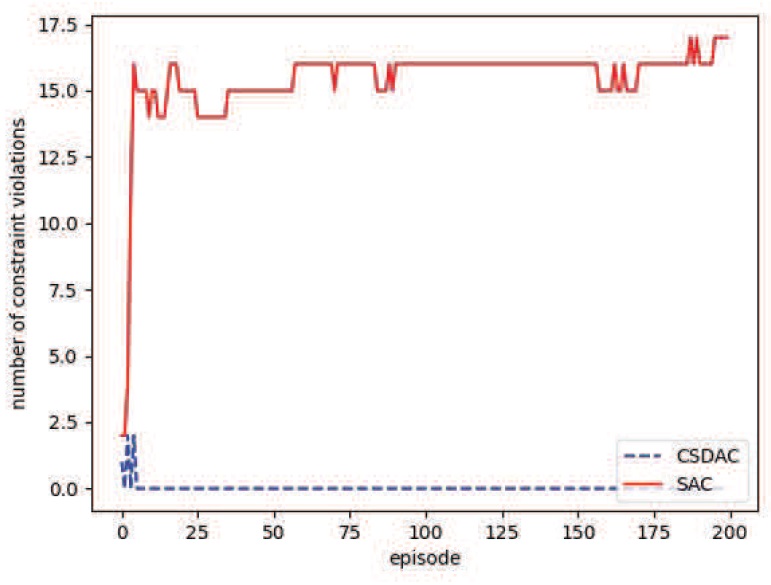}}
	\label{fig:const_break}
	\caption{Number of constraint violations in CSDAC compared to SAC}
\end{figure}
\begin{figure}[h!]
	\centering
	\centerline{\includegraphics[width=0.5\textwidth]{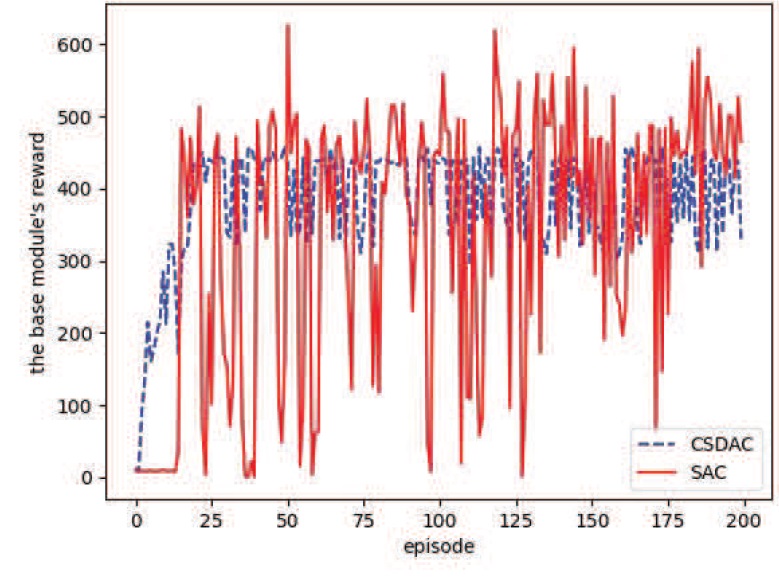}}
	\label{fig:base_reward}
	\caption{The base module's reward trained by CSDAC and SAC}
\end{figure}
\begin{figure}[h!]
	\centering
	\centerline{\includegraphics[width=0.5\textwidth]{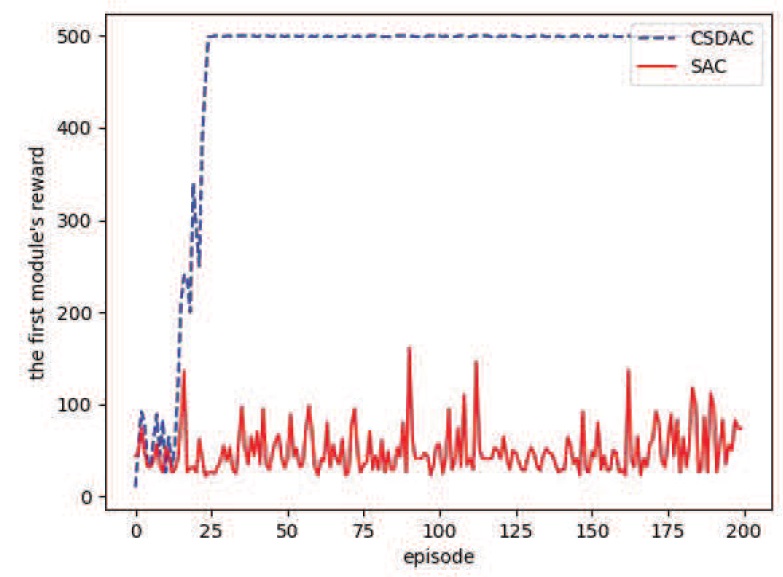}}
	\label{fig:first_module_reward}
	\caption{The first module's reward trained by CSDAC and SAC}
\end{figure}
\section{Conclusion and Future Work}
This paper has concerned with the joint design problem of the event-triggered control and the energy-efficient resource allocation in the 5G-based network. So, a multi-objective optimization problem was formulated to minimize both the number of updates on the actuators' input and the total downlink power usage. The problem constraints contained the dynamics and UUB stability of the control plant, the QoS demands of eMBB and URLLC users, power and sub-carrier constraints, and the BS transmit power limitation. We proposed a novel model-free HRL approach with UUB stability guarantee using CAN method to address the problem. In CAN, the problem was decoupled into two sub-problems. In this regard, we proved that using the decoupling method resulted in a Pareto solution. As the action space of each sub-problem was hybrid, we used DAC architecture to deal with each sub-problem. We demonstrated the effectiveness of the proposed approach by simulation results. Considering joint uplink and downlink transmission will be concentrated on in future work. 
\begin{appendices}
\section{Proof of Lemma\ref{L3}}
\vspace{-1mm}
In the higher-level MDP of the base module, we define:\\
\begin{equation*}
 \quad \quad \quad \mathcal{A}^{H_0}[k]\doteq \{\beta[k] \}\\
\end{equation*}
\begin{equation*}
\mathcal{P}^{H_0} (\mathcal{S}^{H_0}[k+1]|\mathcal{S}^{H_0}[k],\mathcal{A}^{H_0}[k]) \doteq\!\textbf{1}_{\mathcal{A}^{H_0}=\beta[k]}\mathcal{P}^0 (\mathcal{S}^0[k+1]|\mathcal{S}^0[k],\!\beta[k]),\\
\end{equation*}
and
\begin{equation*}
\pi^{H_0} (\mathcal{A}^{H_0}[k] | \mathcal{S}^{H_0}[k]) \doteq \mathcal{P}^0 (\beta[k] | \beta[k-1], s^{c}[k]),
\end{equation*}
therefore:
\begin{equation*}
\mathcal{P}^0(\tau^0 | \pi^0 ,\mathcal{M}^0) = \mathcal{P}^0 (\mathcal{S}^0[0]) \prod_{k= 0}^{K-1} (\mathcal{P}^0(\beta[k] | \mathcal{S}^0[k],\beta[k-1]) \mathcal{P}^0(\mathcal{S}^0[k+1]| \\
\end{equation*}
\begin{equation*}
\mathcal{S}^0[k],\beta[k])) = \mathcal{P}^0 (\mathcal{S}^0[0]) \prod_{k= 0}^{K-1} (\mathcal{P}^0(\beta[k] | S^{0}[k],\beta[k-1])\textbf{1}_{\mathcal{A}^{H_0}=\beta[k]} \\
\end{equation*}
\begin{equation*}
\mathcal{P}^0(S^{0}[k+1]|S^{0}[k],\beta[k])) = \mathcal{P}^{H_0} (S^{H_0}[0]) \prod_{k= 0}^{K-1}(\pi^{H_0} (\mathcal{A}^{H_0}[k] | \mathcal{S}^{H_0}[k])\\ 
\end{equation*}
\begin{equation*}
\mathcal{P}^{H_0}(S^{H_0}[k+1]|S^{H_0}[k],\mathcal{A}^{H_0}[k])) = \mathcal{P}^{H_0}(\tau^{H_0} | \pi^{H_0} ,\mathcal{M}^{H_0}). \qquad \\
\end{equation*}
$\mathcal{R}^0 (\tau^0)= \mathcal{R}^{H_0} (\tau^{H_0})$ follows directly from the definition of $\mathcal{R}^{H_0}$. 
\section{Proof of Lemma\ref{L4}}
\vspace{-1mm}
In the lower-level MDP of the base module, we define:\\
\begin{equation*}
\quad \mathcal{A}^{L_0}[k]\doteq\{u[k] \}\\
\end{equation*}
\begin{equation*}
\mathcal{P}^{L_0} (\mathcal{S}^{L_0}[k+1]|\mathcal{S}^{L_0}[k],\mathcal{A}^{L_0}[k]) \doteq\mathcal{P}^0 ( \beta[k+1],\mathcal{S}^0[k+1] |(\beta[k],\\
\end{equation*}
\begin{equation*}
\mathcal{S}^0[k]),u[k])= \mathcal{P}^0 (\mathcal{S}^0[k+1]|\mathcal{S}^0[k],u[k]) \times\mathcal{P}^0 (\beta[k+1] | \mathcal{S}^0[k+1]\\
\end{equation*}
\begin{equation*}
,\beta[k]),
\end{equation*}
and
\begin{equation*}
\pi^{L_0} (\mathcal{A}^{L_0}[k] | \mathcal{S}^{L_0}[k]) \doteq \mathcal{P}^0 (u[k] | \mathcal{S}^0[k], \beta[k]),
\end{equation*}
therefore:
\begin{equation*}
\mathcal{P}^0(\tau^0 | \pi^0 ,\mathcal{M}^0) = \mathcal{P}^0 (\mathcal{S}^0[0])\mathcal{P}^0 (\beta[0]|\mathcal{S}^0[0]) \prod_{k= 0}^{K-1} (\mathcal{P}^0(u[k] | \mathcal{S}^0[k] \\
\end{equation*}
\begin{equation*}
 ,\beta[k])\mathcal{P}^0(\mathcal{S}^0[k+1]|\mathcal{S}^0[k],u[k]) \mathcal{P}^0(\beta[k+1]|\mathcal{S}^0[k+1],\beta[k]))
\end{equation*}
\begin{equation*}
= \mathcal{P}^{L_0} (S^{L_0}[0]) \prod_{k= 0}^{K-1} (\pi^{L_0} (\mathcal{A}^{L_0}[k] | \mathcal{S}^{L_0}[k]) \mathcal{P}^{L_0}(S^{L_0}[k+1]|S^{L_0}[k], \\
\end{equation*}
\begin{equation*}
\mathcal{A}^{L_0}[k]))= \mathcal{P}^{L_0}(\tau^{L_0} | \pi^{L_0} ,\mathcal{M}^{L_0}). \qquad \qquad \qquad \qquad \qquad \qquad \qquad 
\end{equation*}
$\mathcal{R}^0 (\tau^0)= \mathcal{R}^{L_0} (\tau^{L_0})$ follows directly from the definition of $\mathcal{R}^{L_0}$. 
\end{appendices}
\bibliographystyle{IEEEtran}
	
\vspace{-20pt}
\begin{IEEEbiography}[{\includegraphics[width=1in,height=1.25in,clip,keepaspectratio]{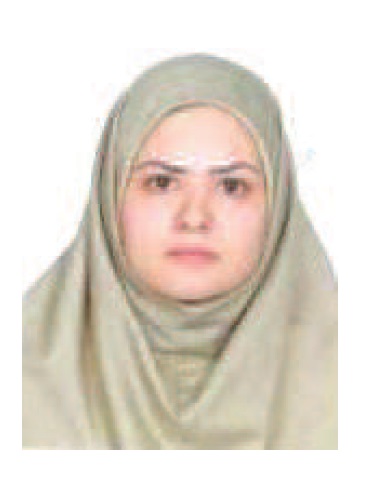}}]
	{Atefeh Termehchi} received her B.Sc. degree in Electrical Engineering from Shiraz University of Technology, Shiraz, Iran, in 2008 and her M.Sc. degree in Electrical Engineering from Amirkabir University of Technology, Tehran, Iran, in 2013. She is pursuing the Ph.D. degree in Information Technology Engineering in Amirkabir University of Technology, Tehran, Iran. Her current research area include reinforcement learning, optimization theory, and their application in industrial cyber physical systems and Beyond 5G wireless networks.
\end{IEEEbiography}
\vspace{-10pt}
\begin{IEEEbiography}[{\includegraphics[width=1in,height=1.25in,clip,keepaspectratio]{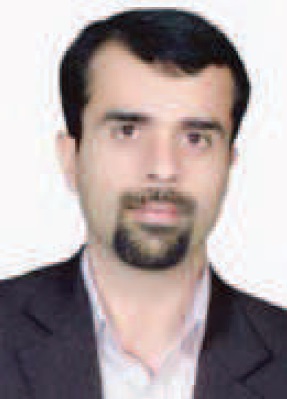}}]
	{Mehdi Rasti} (S'08-M'11-SM'21) is currently an Associated Professor at the Department of Computer Engineering, Amirkabir University of Technology, Tehran, Iran and is a visiting researcher at the Lappeenranta-Lahti University of Technology (LUT), Lappeenranta, Finland. From November 2007 to November 2008, he was a visiting researcher at the Wireless@KTH, Royal Institute of Technology, Stockholm, Sweden. From September 2010 to July 2012 he was with Shiraz University of Technology, Shiraz, Iran. From June 2013 to August 2013, and from July 2014 to August 2014 he was a visiting researcher in the Department of Electrical and Computer Engineering, University of Manitoba, Winnipeg, MB, Canada. He received his B.Sc. degree from Shiraz University, Shiraz, Iran, and the M.Sc. and Ph.D. degrees both from Tarbiat Modares University, Tehran, Iran, all in Electrical Engineering in 2001, 2003 and 2009, respectively. His current research interests include radio resource allocation in IoT, Beyond 5G and 6G wireless networks.
\end{IEEEbiography}
\end{document}